\documentclass[fleqn,usenatbib]{mnras}

\usepackage{newtxtext,newtxmath}

\usepackage[T1]{fontenc}

\DeclareRobustCommand{\VAN}[3]{#2}
\let\VANthebibliography\thebibliography
\def\thebibliography{\DeclareRobustCommand{\VAN}[3]{##3}\VANthebibliography}

\usepackage{graphicx}	
\usepackage{amsmath}	
\usepackage{graphicx}
\usepackage{subcaption}
\usepackage{comment}
\usepackage{pdflscape}	
\usepackage{physics}

\title[The fraction of clumpy star-forming galaxies in nearby galaxies from CLAUDS and HSC-SSP]{The fraction of clumpy star-forming galaxies in nearby galaxies from CLAUDS and HSC-SSP}

\author[J. J. Popp et al.]{
Jürgen J. Popp,$^{1}$\thanks{E-mail: jurgen.popp@open.ac.uk}
Hugh Dickinson,$^{1}$
Stephen Serjeant,$^{1}$
Lucy F. Fortson,$^{2}$
Tobias Géron$^{3}$
\newauthor
and Vihang Mehta$^{4}$
\\
$^{1}$School of Physical Sciences, The Open University, Milton Keynes, MK7 6AA, UK\\
$^{2}$School of Physics and Astronomy, University of Minnesota, 116 Church Street SE, Minneapolis, MN 55455, USA\\
$^{3}$Dunlap Institute for Astronomy and Astrophysics, University of Toronto, 50 St. George Street, Toronto, ON M5S 3H4, Canada\\
$^{4}$IPAC, Mail Code 314-6, California Institute of Technology, 1200 E. California Blvd., Pasadena, CA, 91125, USA\\
}

\date{Accepted XXX. Received YYY; in original form ZZZ}

\pubyear{\the\year{}}

\begin{document}
\label{firstpage}
\pagerange{\pageref{firstpage}--\pageref{lastpage}}
\maketitle

\begin{abstract}
Massive, star-forming clumps are regions of intensive star-formation that are commonly observed in high-redshift ($z > 1$) galaxies. Observations of low-redshift clumpy galaxy analogues are rare but the availability of wide-field galaxy survey data makes the detection of large clumpy galaxy samples much more feasible. We present a population of 12,790 star-forming clumps detected in a mass-complete sample of 5,395 star-forming galaxies (SFGs) at redshifts $z\leq0.32$, located in the XMM-LSS, E-COSMOS and DEEP2-3 fields observed by the Hyper Suprime-Cam Subaru Strategic Survey (HSC-SSP) and CFHT Large Area U-band Deep Survey (CLAUDS). The clumps were detected using an improved version of our Deep Learning (DL)-based object detection framework which uses the \textsc{Zoobot} foundation DL-model as a `backbone' feature extractor. We determined the fraction of star-forming galaxies hosting at least one off-centre clump ($f_{\mathrm{clumpy}}$) based on a clump definition that requires a clump-galaxy flux ratio in the CLAUDS \textit{u}-band of $\geq8\%$. We estimate $f_{\mathrm{clumpy}}$ to decrease from $\sim$31\% at $z\sim0.3$ to $\sim$23\% at $z \sim 0.1$, which aligns well with a low-redshift extrapolation of the clumpy fraction that is measured using high-redshift observations. At fixed redshift, $f_{\mathrm{clumpy}}$ is negatively correlated with the stellar mass and positively correlated with the specific star-formation rate (sSFR) of the host galaxies. When the clump definition is changed to include only clumps with a stellar mass of $M_{\mathrm{cl}} \geq 10^7 M_\odot$, we observe a highly increased clumpy fraction of $\sim$60\% that tends to increase with the stellar mass of the host galaxies but does not show a dependence on the sSFR of the host galaxies.
\end{abstract}

\begin{keywords}
galaxies: star formation -- galaxies: evolution -- galaxies: statistics -- methods: data analysis -- techniques: photometric.
\end{keywords}



\section{Introduction}\label{sec:introduction}
Star-forming galaxies (SFGs) at redshifts $z>1$ often show morphologies that differ from the smooth disk morphologies that are mainly seen for their low-redshift counterparts \citep{Cowie1995,Bergh1996,Conselice2004,Elmegreen2005a,Elmegreen2007a,Elmegreen2009,FoersterSchreiber2009,FoersterSchreiber2011}. In particular, morphological substructures with high surface brightness that were observed in rest-frame ultraviolet (UV) and optical images revealed compact regions of enhanced star-formation within galaxies \citep[e.g.][]{Guo2012,Guo2015,Guo2018,Wuyts2012,Wuyts2013,Shibuya2016,Soto2017}. These star-forming clumps (or `clumps' for short) appear to be much larger and brighter than typical star-forming regions in local SFGs with estimated stellar masses of $10^7$ to $10^9\,M_\odot$ \citep[e.g.][]{FoersterSchreiber2011a,Guo2012,DessaugesZavadsky2017,Soto2017,Guo2018,Zanella2019,HuertasCompany2020,Kalita2025}. The star-formation rates (SFRs) of the clumps range from $10^{-4}$ to $10^2\,M_\odot \,\mathrm{yr}^{-1}$ \citep[e.g.][]{Genzel2011,Guo2012,Elmegreen2013,Guo2018,Soto2017} resulting in specific star-formation rates (sSFRs) that are several times higher than their neighbouring areas in the galaxy.

The formation and evolution modes of star-forming clumps observed in high-redshift galaxies are still debated in the literature. Two principal modes of clump formation have been proposed. The first is the formation by gravitational or violent disk instabilities \citep[VDI,][]{Dekel2009,Dekel2013a} in a gas-rich disk \citep{Elmegreen2005,Bournaud2007,Dekel2009a,Genzel2011,Bournaud2013,Mandelker2014,Romeo2014,Guo2012,Guo2015,Fisher2017,HuertasCompany2020,Claeyssens2025}. The constant supply of accreted cold gas from the intergalactic medium (IGM) provides a high gas-fraction in the disks so that the instabilities are sustained in high-redshift galaxies \citep[e.g.][]{Tacconi2020}. This can then lead to the fragmentation of the gaseous disk and the formation of clumps (in situ clumps). The second mode of formation is believed to be triggered by galaxy-galaxy interactions or minor mergers \citep{Conselice2009,Hopkins2013,Wuyts2014,Ribeiro2017,Mandelker2016,Zanella2019}. Gravitational interactions between galaxies compress the present gas in the host galaxy and initiate strong star-forming activity and/or remnants of accreted smaller galaxies can themselves survive as clumps or clump-like objects (ex situ clumps). Most recent studies suggest that the majority of observed star-forming clumps form in situ as a result of the VDI and only a minor fraction originate from mergers \citep[e.g.][]{Adams2022}. However, the evolution of clumps and their contribution to the evolution of their host galaxy is not yet fully understood.

This contribution is strongly dependent on the longevity of the clumps. Simulation studies show that short-lived clumps ($<100\,\rm{Myr}$) are quickly disrupted by outflows and tidal interactions due to their high sSFR and the resulting strong stellar feedback \citep{Murray2009,Hopkins2012,Hopkins2014,Buck2017,Oklopcic2016}. The disrupted clumps contribute newly formed stars to the formation of thick disks but have only a very weak effect on the stellar structure of the host galaxy otherwise \citep{Genzel2008,Newman2012a}.

Other simulations indicate that clumps survive at least a few orbital timescales and dynamical friction would lead them to migrate towards the galactic centre where they contribute to the growth of the galactic bulge \citep[e.g.][]{Bournaud2007,Elmegreen2008,Ceverino2010,Bournaud2013,Mandelker2014}. This scenario finds support in observations of radial colour gradients of clumps that show redder colours closer to the galactic centre and bluer colours further away \citep[e.g.][]{FoersterSchreiber2011a,Tadaki2014,Shibuya2014,Soto2017,Guo2018}. The colour gradients are also expected to be reflected by decreasing stellar ages with increasing galactocentric distance of the clumps with age differences between the outer and inner clumps that are predicted to be of a few hundred Myr \citep{Dekel2009,Ceverino2010,Dekel2022}. 

The number of SFGs that have at least one off-centre clump in relation to the total number of galaxies defines the clumpy fraction $f_{\mathrm{clumpy}}$. The clumpy fraction, determined over various redshift intervals, likely traces the physical conditions (e.g. gas accretion rate, gas fraction and star-formation efficiency) that drive the formation and evolution of galaxies \citep[e.g.][]{Guo2015,Shibuya2014} and many studies to date have reported clumpy fractions for high-redshift galaxies \citep[for a recent list of studies on clumpy galaxies see][]{Sattari2023,Vega2025}.

The clumpy fraction is determined for a sample $N$ of SFGs, such that:
\begin{equation}\label{eq:gsfc_fclumpy}
    f_{\mathrm{clumpy}} = \frac{N(\mathrm{SFGs\geq 1\,\mathrm{clump}})}{N(\mathrm{SFGs})}.
\end{equation}

Various studies identified an evolutionary trend of the clumpy fraction which increases from $z\sim 8$ to $z \sim 1$-3 and starts to decrease from $z\lesssim 1$ \citep{Murata2014,HuertasCompany2020,Sok2022,Mercier2026}. The peak of the clumpy fraction ($f_{\mathrm{clumpy}}\simeq 60\%$) at redshift $z\simeq 1$-3 \citep[e.g.][]{Cowie1995,Bergh1996,Elmegreen2004,Elmegreen2005,Elmegreen2009,Guo2015,Shibuya2016,Sattari2023,Vega2025} is coincident with the peak of the cosmic star formation density \citep{Madau2014}, but the reported values differ significantly between $\sim$30\% and $\sim$80\% (see Fig. \ref{fig:hsc_phys_props_galaxy_clumpy_fraction_high_z}, for example).

The measured clumpy fractions disagree even more for redshift $z>3$. While the previous observations with the Hubble Space Telescope (HST) indicate that the evolution of galaxies with irregular and clumpy morphologies towards regular disk morphologies happened late in the cosmological timeline, studies using data from JWST have revealed a more complex picture of the clumps and their host galaxies. A significant population of clumps that was not detected by the HST was reported by \citet{Claeyssens2023} and \citet{Fujimoto2025} from gravitationally lensed galaxies over a redshift range of $1<z<8.5$. Additionally, recent JWST observations from unlensed galaxies at $z>4$ also revealed multiple star-forming clumps \citep{Tacchella2023,Tanaka2024,Bik2024,Hainline2024,Vega2025}, indicating that the abundance of clumpy galaxies is higher than previously concluded from pre-JWST observations.

Comparisons between measurements of the clumpy fraction $f_{\mathrm{clumpy}}$ are complicated as the different studies use different clump definitions and different clump detection methods. For example, \citet{Guo2018} and \citet{Adams2022} define star-forming clumps based either on the UV luminosity ratio or \textit{u}-band flux ratio of the clump to its host galaxy. This definition is based on the empirical studies by \citet{Guo2015} and similar definitions have been applied by other clump studies since then \citep[e.g.][]{Shibuya2016,Mandelker2016,Fisher2017a}. In contrast, \citet{HuertasCompany2020} require their star-forming clumps to be more massive than $10^7\,M_\odot$ and other authors do not apply any selection cut based on flux ratios or physical properties of the clumps \citep[e.g.][]{Sattari2023,Claeyssens2025,Vega2025}.

To detect and select star-forming clumps in small samples of a few hundred low- or high-redshift galaxies most studies have applied different transformations of the observed imaging data to obtain high contrast images that emphasise regions in the target galaxy with locally increased surface brightness. These regions have then been identified using source extraction or peak finding algorithms \citep[e.g.][]{Guo2015,Fisher2017,Mehta2021,Lenkic2021,Mestric2022,Sattari2023,Kalita2024,Sok2025a,Claeyssens2025,Vega2025}. Other studies have relied on visual identification by experts \citep[e.g.][]{Elmegreen2007a,Overzier2009,Claeyssens2023}, which is limited to even smaller galaxy samples. Only a few recent studies have analysed clump detections in samples that contain $>1500$ galaxies, either with the help of volunteers participating in citizen science projects like the `Galaxy Zoo: Clump Scout' project \citep[GZCS,][]{Adams2022,Dickinson2022} or by applying Machine Learning (ML) methods to extract the clump locations from the target galaxies \citep{HuertasCompany2020,Popp2024,Adams2025}.

So far, only a few studies have observed star-forming clumps in low-redshift or nearby galaxies \citep[e.g.][]{Overzier2009,Fisher2014,Messa2019,Mehta2021,Lenkic2021,Adams2022} and measurements of $f_{\mathrm{clumpy}}$ for large samples of SFGs at $z<0.5$, to better constrain the evolutionary trend of the clumpy fraction for a continuous redshift range between $0 \leq z \lesssim 0.5$, are still missing. This is partly due to the limited availability of high-resolution imaging data from the Hubble Space Telescope (HST) and the James Webb Space Telescope (JWST) for high-redshift galaxies. Such studies are also made more difficult by the apparent scarcity of low-redshift galaxies that host plausible analogues of the observed clumps at higher redshift.

In this paper, we use data from the Canada-France-Hawaii Telescope (CHFT) Large Area U-bands Deep Survey \citep[CLAUDS,][]{Sawicki2019} and the Hyper-Suprime-Cam (HSC) Subaru Strategic Program \citep[HSC-SSP,][]{Aihara2017} to identify star-forming clumps in low-redshift galaxies using an object detection model that is based on Deep Learning (DL) techniques \citep{Popp2026b}. We estimate $f_{\mathrm{clumpy}}$ for our mass-complete sample of 7,778 galaxies over the redshift range of $0.005<z<0.32$ based on a clump definition that requires a clump-galaxy flux ratio in the CLAUDS \textit{u}-band of $\geq8\%$. Furthermore, we investigate how the clumpy fraction changes with increasing stellar mass and sSFR of the host galaxy and compare how a changed clump definition can lead to different results.

This paper is organised as follows. Section \ref{sec:data} describes our galaxy sample and Section \ref{sec:clump_identification} briefly introduces our DL-based clump detection model. This is then followed by a description of the photometry method that we applied to measure the fluxes of the detected clumps (Section \ref{sec:clump_props}). In Section \ref{sec:clumpy_fraction}, we present our measurements of $f_{\mathrm{clumpy}}$ and outline how we corrected our observed clumpy fraction for incompleteness. We discuss and compare our results to other published work from the literature in Section \ref{sec:discussion} and conclude with a brief summary in Section \ref{sec:conclusion}.

Throughout this paper, we express all magnitudes in the AB system \citep{Oke1983}. For simplicity, we use the terms `low-redshift' for a redshift range of $z\leq 0.5$ and `high-redshift' for a range of $z>0.5$. Logarithmic quantities are either referenced to a base of $10$ using the notation $\log$ or to a base of $e$ using the notation $\ln$. In this work we adopt the Planck 2015 \citep{Ade2016} cosmological parameters with $(\Omega_m, \Omega_{\Lambda}, h) = (0.31, 0.69, 0.68)$.

\section{Data}\label{sec:data}
For our analysis we combined imaging data from CLAUDS \citep{Sawicki2019} and HSC-SSP \citep{Aihara2017}. HSC-SSP is a multiband, three-layered imaging survey that covers $1400\,\mathrm{deg}^2$ (Wide survey), $27\,\mathrm{deg}^2$ (Deep survey) and $3.5\,\mathrm{deg}^2$ (Ultra-Deep survey) of the sky. Observations are made with the HSC on the $8.2\,\mathrm{m}$ Subaru Telescope using five broadband (\textit{grizy}) and additional narrowband filters with $5\sigma$ point source depths in the \textit{r}-band of $\sim 26\,m_{\mathrm{AB}}$, $\sim 27\,m_{\mathrm{AB}}$ and $\sim 28\,m_{\mathrm{AB}}$ for the Wide, Deep and Ultra-Deep survey, respectively. CLAUDS provides \textit{u}-band imaging data for those areas where it overlaps with the HSC-SSP survey for the XMM-LSS, E-COSMOS, ELAIS-N1 and DEEP2-3 fields. The \textit{u}-band imaging data was acquired using two different filters on the CFHT MegaCam. The older $u^\star$ filter was mainly used for the XMM-LSS field and later replaced by the $u$ filter which was used for the E-COSMOS, ELAIS-N1 and DEEP2-3 fields. The two different \textit{u}-band filters cover slightly different wavelength ranges and are treated differently throughout image processing and photometry related tasks (Tab. \ref{tab:hsc_data_survey_overview}).

The HSC-SSP imaging data were processed by the HSC pipeline \citep[HSCpipe,][]{Bosch2017}, which was also used to process the CLAUDS images \citep{Sawicki2019}. Apart from the imaging data, both surveys provide catalogues with multiple photometry measurements in the different filter bands and inferred physical properties of the detected source objects, which are described in \citet{Desprez2023} for the CLAUDS catalogue and in \citet{Aihara2022} for the HSC-SSP Public Data Release 3 (PDR3) catalogue. For consistency, we use and report the published source properties from the HSC-SSP PDR3 catalogue as they are provided for all objects in our selected sample. All object identifiers used in this work refer to the object IDs in the HSC-SSP PDR3 catalogue.

\subsection{Galaxy sample}\label{sec:data_gal_sample}
The selection of our target galaxies is described in detail in \citet{Popp2026b}. Briefly, we started with a preselection of target galaxies from the Sloan Digital Sky Survey (SDSS) Data Release 18 \citep[SDSS DR18,][]{Almeida2023} catalogue. By doing so we obtain an initial sample of galaxies with robust measurements of the Petrosian radius that we used to define the cutout size to ensure a comparable visual size of the target galaxies in each cutout. We also required the galaxies to have a minimum extent defined by a SDSS \textit{r}-band 90\% Petrosian radius of $\geq3\,\mathrm{arcsec}$ so that morphological features are resolvable. This preselection was then crossmatched (within 1.0 arcsec) with sources detected in the \textit{i}-band from the latest Public Data Release 3 \citep[PDR3,][]{Aihara2022} of the HSC-SSP Wide survey, which resulted in 710,271 HSC-SSP galaxies with clean five filter band photometry (\textit{grizy}, Table \ref{tab:hsc_data_preprocess}). We then crossmatched the HSC-SSP set with the CLAUDS data for the XMM-LSS, E-COSMOS and DEEP2-3 fields. The ELAIS-N1 field is covered by the HSC-SSP Deep and Ultra-Deep surveys but not by the HSC-SSP Wide survey (using the same crossmatching distance of 1.0 arcsec) so that our final set consists of 14,231 galaxies with six filter band photometry data (\textit{ugrizy}, Table \ref{tab:hsc_data_preprocess}). We chose to use the data from the HSC-SSP Wide survey instead of the Deep and Ultra-Deep surveys because this allowed us to extend our clump analysis to the much wider sky area of the HSC-SSP Wide survey \citep[see also][and Popp et al. 2026c, submitted]{Popp2026b}. The final set is magnitude-limited by the median $5\sigma$ depth for SDSS photometric observations in the $r_{\mathrm{SDSS}}$-band magnitude of $r_{\mathrm{SDSS}} \leq 22.7\,m_{\mathrm{AB}}$ and limited to a maximum redshift of $z \leq 0.5$.

We downloaded image cutouts for each target galaxy from the HSC-SSP Data Archive System (DAS) using the provided web-based interface\footnote{\url{https://hsc-release.mtk.nao.ac.jp/das_cutout/pdr3/}}. The cutouts were centred on the RA/DEC position of the target galaxy as reported from the HSC-SSP PDR3 catalogue and the cutout size set to a square with an edge length of twice the SDSS r-band 90\% Petrosian radius in arcsec ($2 \times \texttt{petroR90\_r}$). For each target galaxy, we downloaded cutouts for the \textit{g}-, \textit{r}-, \textit{i}-, \textit{z}- and \textit{y}-filter band as separate files containing the sky-subtracted and calibrated science image and the corresponding variance map. 

\textit{U}-band image cutouts from the CLAUDS data were made at the same RA/DEC positions as the crossmatched HSC-SSP galaxies and with the same cutout size. The CLAUDS imaging data is only available as co-added stacks of sky-subtracted and calibrated tiles that cover $4200\times 4100$ pixel each or $\sim 0.2\times 0.2\,\mathrm{deg}$ (pixel scale of $\sim 0.168\,\mathrm{px}/\mathrm{arcsec}$) and the science cutouts were generated from the tiles using functions provided by the Python library \textsc{Astropy} \citep{astropy2022}.

Even though CLAUDS was designed to closely match and overlap with the HSC-SSP surveys, some crossmatched target galaxies that are located close to the CLAUDS survey field borders were only partially imaged by CLAUDS. The target galaxies for which the \textit{u}-band cutouts contain only partial data were discarded. Furthermore, weight/variance maps were not available for some target galaxies and these were also excluded from the final galaxy sample. We also excluded images of target galaxies that either have too many faulty or unreasonable pixel values. Table \ref{tab:hsc_data_preprocess} lists the number of galaxies at each stage of our selection process.

\begin{table}
	\centering
	\caption[Filter bands, depths and wavelength ranges used by the CLAUDS and HSC-SSP Wide survey.]{Filter bands, depths and wavelength ranges used by the CLAUDS \citep{Sawicki2019} and HSC-SSP Wide survey \citep{Aihara2017}. The CLAUDS \textit{u}-band filter combines the CFHT MegaCam filters $u$ and $u^\star$. The $u$-filter is used for the E-COSMOS, ELAIS-N1 and DEEP2-3 fields while the older $u^\star$-filter only for the XMM-LSS field. The limiting magnitude is shown as the $5\sigma$ point source depth and seeing as the median seeing FWHM with the seeing range of our galaxy sample in brackets.}
    \label{tab:hsc_data_survey_overview}
	\footnotesize
    \begin{tabular}{llrrr}
		\hline
		\multicolumn{1}{c}{Survey} & \multicolumn{1}{c}{Filter-} & \multicolumn{1}{c}{Seeing} & \multicolumn{1}{c}{Lim. mag.} & \multicolumn{1}{c}{Wavelength} \\ 
         & \multicolumn{1}{c}{band} & \multicolumn{1}{c}{[arcsec]} & \multicolumn{1}{c}{$[m_{\mathrm{AB}}]$} & \multicolumn{1}{c}{range, [nm]} \\
		\hline
		CLAUDS   & \textit{u}/$u^\star$ & 0.92        & 27.1 & 310-397 / \\
                 &            & (0.69-1.15) &      & 336-412 \\
        HSC Wide & \textit{g} & 0.79        & 26.5 & 400-500 \\
                 &            & (0.51-1.01) &      & \\
        HSC Wide & \textit{r} & 0.75        & 26.1 & 550-695 \\
                 &            & (0.45-1.03) &      & \\
        HSC Wide & \textit{i} & 0.61        & 25.9 & 695-845 \\
                 &            & (0.41-0.74) &      & \\
        HSC Wide & \textit{z} & 0.68        & 25.1 & 845-930 \\
                 &            & (0.54-0.87) &      & \\
        HSC Wide & \textit{y} & 0.68        & 24.4 & 930-1,070 \\
                 &            & (0.50-1.00) &      & \\
		\hline
	\end{tabular}
\end{table}

\subsection{Mass-completeness of the galaxy sample}\label{sec:data_gal_sample_compl}
To determine the 90\% mass-completeness limit for our sample of galaxies over the observed redshift range, we followed the method described by \citet{Pozzetti2010}. For every galaxy, a stellar mass limit $M_{\star,\mathrm{lim}}$ was calculated that represents the mass a galaxy would have if its apparent magnitude was equal to the SDSS \textit{i}-band detection limit at $i_{\mathrm{lim}} = 22.2\,m_{\mathrm{AB}}$. With the observed stellar mass $M_\star$ of the galaxy, $M_{\star,\mathrm{lim}}$ can be calculated as:
\begin{equation}
    \log(M_{\star,\mathrm{lim}}) = \log(M_\star) + 0.4 (i-i_{\mathrm{lim}}),
\end{equation}
where $i$ is the SDSS \textit{i}-band magnitude of the galaxy.

\begin{figure}
    \centering
    \includegraphics[width=0.4\textwidth]{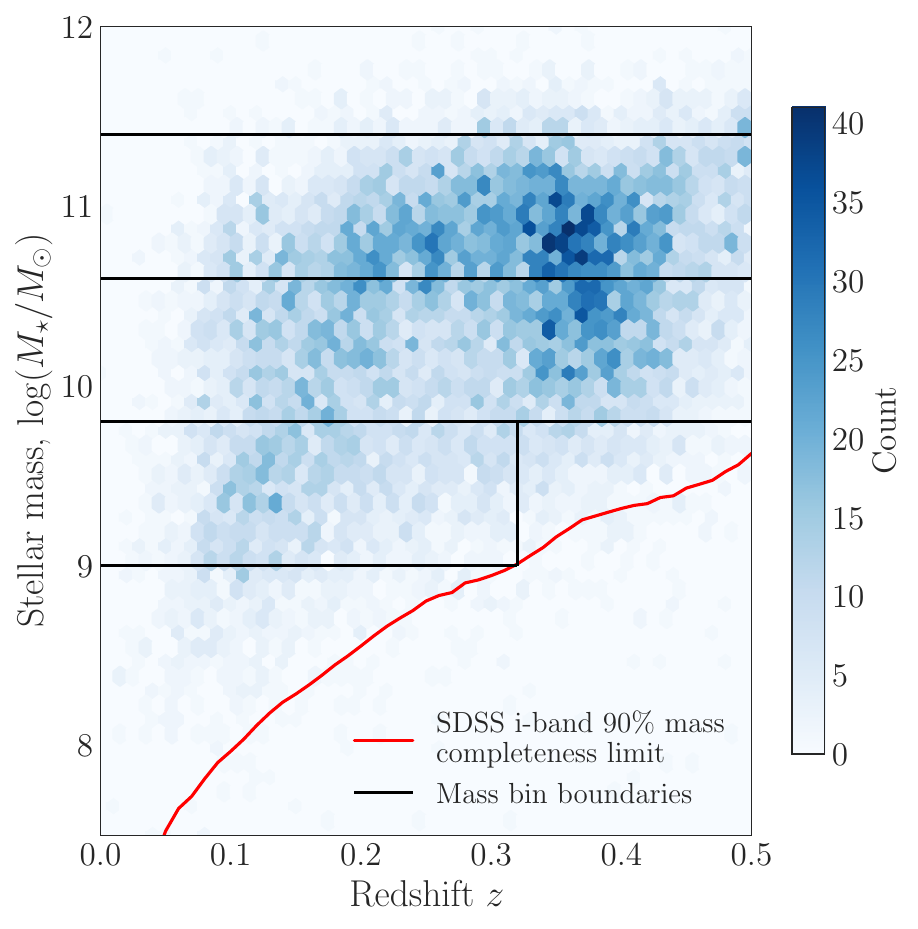}
    \caption[Galaxy stellar mass as a function of redshift for the sample of CLAUDS and HSC-SSP galaxies.]{Galaxy stellar mass as a function of redshift for the sample of CLAUDS and HSC-SSP galaxies. The red dotted line indicates the 90\% mass-completeness limit as a function of redshift. The black outlines mark the mass-complete sample for the mass bins $9.0 \leq \log(M_\star/M_\odot) < 9.8$ (redshift $z<0.32$), $9.8 \leq \log(M_\star/M_\odot) < 10.6$ and $10.6 \leq \log(M_\star/M_\odot) < 11.4$.}
    \label{fig:hsc_data_galaxy_completeness}
\end{figure}

Then, for each redshift bin between $0.0 \leq z \leq 0.5$ using a bin width of $\Delta z = 0.01$, we selected the faintest 20\% of the galaxies and determined the 90th percentile of $M_{\star,\mathrm{lim}}$ for each subset of galaxies. Figure \ref{fig:hsc_data_galaxy_completeness} shows the resulting 90\% mass-completeness limit as a function of redshift for the sample of CLAUDS and HSC-SSP galaxies.

Figure \ref{fig:hsc_data_galaxy_completeness} also shows the three different galaxy stellar mass bins that are used for a comparison of the clumpy fractions in Section \ref{sec:clumpy_fraction}. The galaxy sample is mass-complete for galaxies with stellar mass $\log(M_\star/M_\odot) \geq 9.0$ and redshift $z \leq 0.32$. The mass-complete sample used in this paper includes 7,778 galaxies with six filter band (\textit{ugrizy}) imaging data. The counts for each stellar mass bin are listed in Table \ref{tab:hsc_data_preprocess}.

\begin{table}
	\centering
	\caption[Number of galaxies after each selection and exclusion step.]{Number of galaxies after each selection and exclusion step. The mass-complete sample consists of all galaxies with redshift $z \leq 0.32$ and stellar mass $\log(M_\star/M_\odot) \geq 9.0$. The stellar mass bins for galaxies with redshift $z \leq 0.32$ are: $9.0 \leq \log(M_\star/M_\odot) < 9.8$ (low), $9.8 \leq \log(M_\star/M_\odot) < 10.6$ (medium) and $10.6 \leq \log(M_\star/M_\odot) < 11.4$ (high).}
    \label{tab:hsc_data_preprocess}
	\footnotesize
        \begin{tabular}{lr}
		\hline
		Selection & \multicolumn{1}{c}{Galaxy count} \\
		\hline
        SDSS DR18 preselection                 & 11,725,567 \\
        Crossmatches with HSC-SSP PDR3         & 1,357,190 \\
        With clean photometry                  & 710,271 \\
        \hline
        Crossmatches with CLAUDS               & 19,811 \\
        With science image                     & 15,952 \\
        With weight/variance map               & 15,934 \\
        \hline
        After faulty image map exclusions      & 14,231 \\ 
        \textit{thereof:} mass-complete        &  7,778 \\
        \textit{thereof:} low-mass galaxies    &  2,073 \\
        \textit{thereof:} medium-mass galaxies &  3,002 \\
        \textit{thereof:} high-mass galaxies   &  2,703 \\
		\hline
	\end{tabular}
\end{table}

\section{Identifying star-forming clumps}\label{sec:clump_identification}
We identified potential star-forming clumps or clump candidates in our galaxy sample using a DL-based object detection model that is described in detail in \citet{Popp2026b}. For convenience, we describe our model development process briefly in Appendix \ref{sec:frcnn_development} and summarise only the main steps in the following paragraphs.

The model is built on the Faster R-CNN object detection framework \citep[FRCNN,][]{Ren2015} but accepts six channel data as input instead of the three channels (e.g. RGB) of its basic version. As its feature extracting backbone, it uses the \textsc{Zoobot} foundation deep learning model \citep{Walmsley2023} in its latest version v2.0. Such a domain-specific convolutional neural network (CNN) that has been pretrained for image classification using large sets of astrophysical images improves the detection performance while the FRCNN model needs to be trained on only a relatively small sample of labelled data for the specific downstream task \citep{Popp2024}. Our trained FRCNN model outputs bounding boxes around detected clump candidates and possible contaminating objects and applies a classification scheme with multiple classes that include object classes for clumps, foreground stars, fore-/background galaxies, bulges and the generic background.

\begin{figure*}
    \centering
    \includegraphics[width=1.0\textwidth]{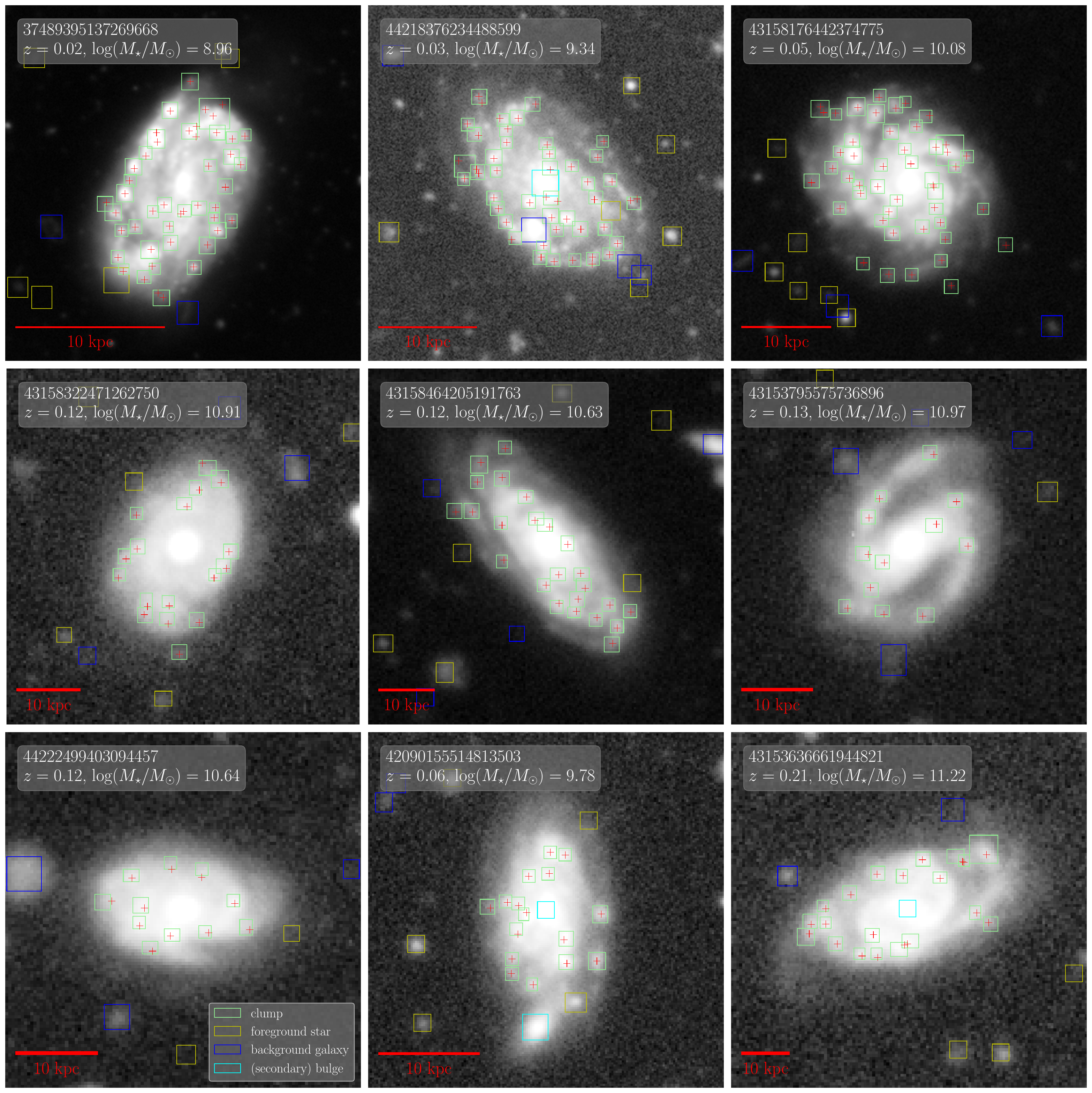}
    \caption[Galaxy examples showing the postprocessed FRCNN model detections with extracted flux peaks.]{Galaxy examples showing the postprocessed FRCNN model detections with extracted flux peaks. The model detections are shown as boxes where the colour indicates the object class. Flux peaks are marked with red crosses. The galaxies are shown with their \textit{u}-band images.}
    \label{fig:hsc_det_over_peaks_example}
\end{figure*}

We applied the object detection model and postprocessed the detection results as described in \citet[][see also Appendix \ref{sec:frcnn_development}]{Popp2026b}. First, we constructed greyscale images for each of the six \textit{ugrizy}-filter band science images of our sample of CLAUDS/HSC-SSP galaxies that we also resized to $400\times 400$ pixels. After the object detection model was applied to all images, the detection results were postprocessed by (1) using non-maximum suppression (NMS), (2) removing bounding boxes that are larger than the 95th percentile threshold of the bounding box size distribution (corresponding to a maximum bounding box size of $\leq 7.30\,\mathrm{kpc}$), (3) removing clump detections for which the centroid of the bounding boxes lie outside the target galaxy's segmentation mask (see Appendix \ref{sec:hsc_data_gal_extent}) and (4) by discarding those bounding boxes that are close to or coincide with the central bulge of a galaxy. Here, a bounding box is considered to be marking the centre of a galaxy instead of an off-centre feature if the distance of its midpoint to the flux-weighted centroid of the galaxy is less than 2\% of the Petrosian radius, which we have remeasured for each galaxy (see Appendix \ref{sec:data_preprocess_reff}). As a final step, we discarded all bounding boxes that the FRCNN model classified as contaminating objects (i.e. non-clumps, see also Appendix \ref{sec:frcnn_development}) and extracted the local flux maxima or flux peaks within each remaining clump bounding box. These peaks are marking the positions of our final sample of clump candidates. In Figure \ref{fig:hsc_det_over_peaks_example}, we show the bounding boxes of the model detections together with the extracted flux peaks for nine galaxy examples. In total, we identified 30,148 clump bounding boxes in 7,135 galaxies and extracted 30,636 clump candidates.

\section{Clump photometry}\label{sec:clump_props}

\subsection{PSF estimation and image convolution}\label{sec:phot_psf}
We convolved all six \textit{ugrizy} science images of a galaxy cutout with a kernel that is generated from the point-spread function (PSF) model corresponding to the band-specific image and a reference PSF model. The reference PSF is the PSF from the image with the largest seeing full width at half maximum (FWHM) over the \textit{ugrizy}-filter bands for each object as reported from the observation metadata (Table \ref{tab:hsc_data_survey_overview}). The HSC-SSP PDR3 catalogue provides discretised PSF models at the sky location for all detected objects in the survey catalogue \citep[][accessible through the HSC PSF picker tool\footnote{\url{https://hsc-release.mtk.nao.ac.jp/psf/pdr3/}}]{Aihara2022}. A similar service for the \textit{u}-band images from CLAUDS does not exist. Instead, we constructed an effective PSF (ePSF) model for the \textit{u}-band images by combining PSF estimates based on bright stars selected from the CLAUDS source catalogue that are located in the vicinity of our target galaxies. As the \textit{u}-band galaxy stamps were too small to contain enough bright stars, we used the full image patches, from which the stamps were cut. For each patch with an extent of $4200 \times 4100$ pixels, we selected all stars with an apparent \textit{u}-band magnitude $u \leq 24.0\,m_{\mathrm{AB}}$ and for which no detection error flags were reported.Around 80 bright stars were used to build an ePSF for each of the 750 patches.

Each selected star was extracted as a $43 \times 43$ pixel cutout from which the local background around the source object was estimated and subtracted, and its value used as input for the \texttt{EPSFBuilder} class from \textsc{Photutils} \citep{LarryBradley2025}. This class provides functions to estimate the effective PSF following the algorithm described by \citet{Anderson2000}. The pixel grid was sampled with an oversampling factor of 1.0 (i.e. no oversampling) where the seeing FWHM is $\geq 3$ pixels (or $\sim 0.5$ arcsec) and a factor of 2.0 where the seeing FWHM is below 3 pixels \citep[similar to the HSC PSF models,][]{Bosch2017}. We then assigned each \textit{u}-band galaxy image the ePSF from the patch the galaxy is contained in.

\subsection{Aperture photometry}\label{sec:clump_phot}
We measured the \textit{ugrizy}-fluxes of the 30,636 clump candidates at their positions in the \textit{u}-band images that were determined by extracting one or more flux peaks within the bounding boxes of the model detections (Section \ref{sec:clump_identification}). These positions were held fixed for the other \textit{g}-, \textit{r}-, \textit{i}-, \textit{z}- and \textit{y}-band images (i.e. we extracted forced photometry for those bands). We further assumed that the clumps are unresolved sources and can be treated as point-like objects over the redshift range of our sample of target galaxies given the average CLAUDS \textit{u}-band seeing of $\sim0.9\,\mathrm{arcsec}$.

The flux of each clump candidate was measured using an aperture centred on the clump's position. We estimated the diffuse galaxy background light from an annulus around the aperture and subtracted the median value per pixel from the aperture pixel values. However, the main challenges for measuring the fluxes of our sample of clump candidates lie in excluding contamination from neighbouring objects and the underlying host galaxy light as accurately as possible. We therefore masked the area outside the host galaxy extent (see Appendix \ref{sec:hsc_data_gal_extent}) and adjacent clump detections as their light would otherwise contaminate the background estimate if they are (partly) located within the annulus. The adjacent clump locations were masked out with a circular mask. The radius of the adjacent clumps' masks controls their influence on the background estimate and has, together with the size of the aperture and the size of the annulus, a significant effect on the accuracy of the flux measurements of the clumps. 

To account for the varying imaging data quality of each galaxy cutout, we set the sizes for the aperture radius, annulus radii and clump mask radius as multiples of the filter band-specific seeing FWHM (Table \ref{tab:hsc_data_survey_overview}). We further corrected the measured flux for the flux that is not included within the aperture because of the extended shape of the PSF using the two-dimensional PSF models (Section \ref{sec:phot_psf}) for each filter band and galaxy image over the area of the aperture. The sum of the pixel-values enclosed by the aperture for the model PSF divided by its total pixel sum is used as the aperture correction factor.

We tested a range of different values for all three parameters of our aperture photometry setup on a subset of 605 galaxies that contain clump bounding boxes with multiple clumps (or flux peaks). These galaxies mostly show a complex background with many clumps close to each other and are therefore examples where different radii for the apertures, the annuli and clump masks will have a strong effect. For the test, we used simulated clumps that were injected into the galaxy images and used to validate the performance of the clump detection model from \citet[][see also Appendix \ref{sec:frcnn_development}]{Popp2026b}. We compared the recovered photometry measurements of 3,449 simulated clumps in the galaxy subset to their true values and also included the 5,046 real clump candidates detected by the FRCNN model close to the simulated clumps when masking the annuli for background estimation.

\begin{table}
	\centering
	\caption[Parameters and radii values for testing the aperture photometry of clumps.]{Parameters and radii values for testing the aperture photometry of clumps. The values are given in units of the seeing FWHM, which is specific to each filter band (see Table \ref{tab:hsc_data_survey_overview})}
    \label{tab:hsc_phot_overview_aper_apertest} 
	\footnotesize
    \begin{tabular}{lll}
	\hline
	Parameter & Values & Final value\\ 
	\hline
	Aperture     & $0.125, 0.250, 0.375, 0.500, 0.625,$ & $0.250$ \\
                 & $0.750, 0.875, 1.000$ & \\
    Annulus      & $(1.0, 1.25), (1.0, 1.5), (1.0, 2.0),$ & $(1.5, 2.0)$ \\
    (min., max.) & $(1.25, 1.5), (1.25, 2.0), (1.5, 2.0),$ & \\
                 & $(1.75, 2.0)$ & \\
    Clump mask   & $0.5, 0.75, 1.0, 1.25, 1.5, 1.75, 2.0$ & $0.5$ \\
	\hline
	\end{tabular}
\end{table}

In total, we tested $392$ different combinations of all three parameters with values shown in Table \ref{tab:hsc_phot_overview_aper_apertest}. We found that the lowest median difference between the recovered and the true flux values over all six filter bands combined is achieved with an aperture radius of $r_{\mathrm{ap}}=0.25\, \mathrm{FWHM}_{\mathrm{seeing}}$, an annulus spanning a radial range from $r_{\mathrm{an}}=1.5-2.0\, \mathrm{FWHM}_{\mathrm{seeing}}$ and a clump mask with radius $r_{\mathrm{mask}}=0.5\, \mathrm{FWHM}_{\mathrm{seeing}}$ (see also Table \ref{tab:hsc_phot_overview_aper_apertest}). 

\begin{figure}
    \centering
    \includegraphics[width=0.4\textwidth]{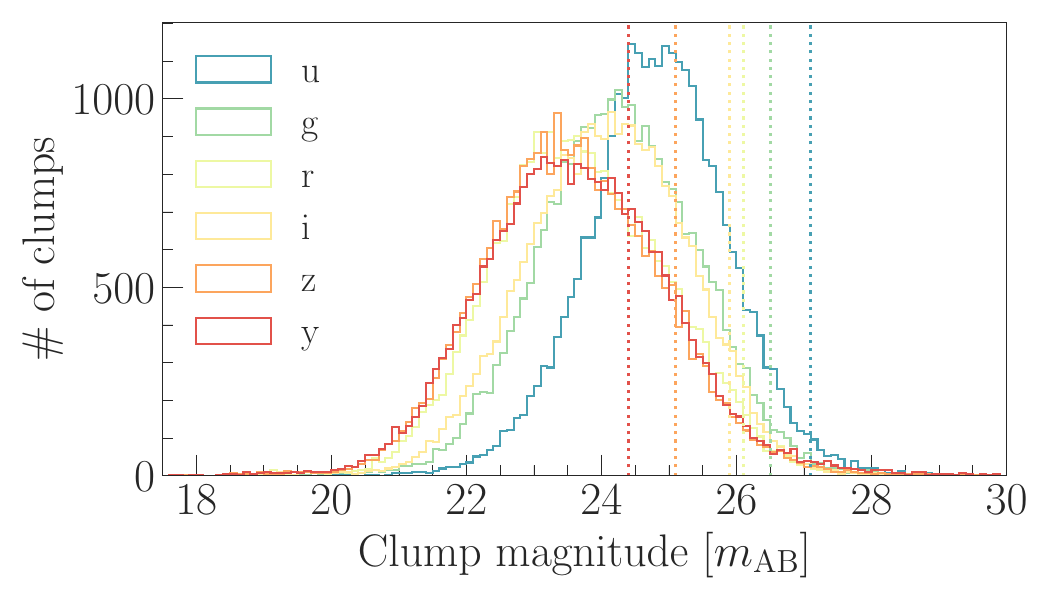}
    \caption[Distribution of the apparent magnitudes of the detected clump candidates.]{Distribution of the apparent magnitudes of the detected clump candidates for each filter band. The coloured dotted lines mark the detection limit for each filter band.}
    \label{fig:hsc_phot_final_mag_distribution_UGRIZY}
\end{figure}

From the 30,636 clump candidates, a valid flux value was measured for 28,814 clump candidates but for 1,822 clumps, or $5.95\%$, the measurement returned either a negative flux value or the image contained missing per-pixel flux values at or around the pixel location of the measurement. The valid flux measurements were converted into AB-magnitudes (see the distributions in Figure \ref{fig:hsc_phot_final_mag_distribution_UGRIZY}) and corrected for Galactic extinction. The reddening $E(B-V)$ is calculated using the \citet{Schlegel1998} dust maps with an extinction to reddening ratio of $R_{V}=3.1$ for the Milky Way.

Of the 28,814 clump candidates with valid flux measurements, 9,050 ($31.4\%$) were fainter than the filter band-specific detection limits in at least one filter band. The remaining 19,764 ($68.6\%$) clump candidates have measured magnitudes of $m_{u}\leq 27.1\,m_{\mathrm{AB}}$, $m_{g}\leq 26.5\,m_{\mathrm{AB}}$, $m_{r}\leq 26.1\,m_{\mathrm{AB}}$, $m_{i}\leq 25.9\,m_{\mathrm{AB}}$, $m_{z}\leq 25.1\,m_{\mathrm{AB}}$ and $m_{y}\leq 24.4\,m_{\mathrm{AB}}$.

\subsection{Validation of the photometry measurements}\label{sec:clump_phot_val}
To validate the photometry measurements, we used the set of simulated clumps \citep[see Appendix \ref{sec:frcnn_development} and][]{Popp2026b} and a sample of star-like objects from the HSC-SSP/CLAUDS catalogue. After running our model on the galaxy images with simulated clumps injected, its detections were then crossmatched with the ground-truth set of simulated clumps, for which the coordinates are known. We counted a successful detection if the distance between the extracted flux peak of a predicted simulated clump and a `true' simulated clump is less than 0.75 of the image-specific \textit{u}-band seeing FWHM. Of the 32,241 simulated clumps, 7,161 were detected and correctly classified as a clump by the FRCNN model (but see Section \ref{sec:clumpy_fraction_incomplete} for the increased completeness after applying a clump-galaxy \textit{u}-band flux ratio threshold $F_{u,\mathrm{cl}}/F_{u,\mathrm{gal}} \geq 0.08$). 

The star-like catalogue objects were selected by taking a subset of 514 galaxy cutouts (500 randomly selected cutouts and 14 manually selected cutouts, for which we expected complex backgrounds that would be hard to measure and correct for, and querying the HSC database for non-extended objects located within each cutout's extent. Non-extended objects were selected by requiring the I-band extendedness flag to be \texttt{i\_extendedness\_value} $< 0.5$ and excluding objects with bad or unreliable photometry. Crossmatching with the CLAUDS catalogue (matching distance $d \leq 1$ arcsec) yielded a set of 1,178 objects in 296 cutouts, of which 611 objects are located within the galaxy extent. These 611 objects form the test set for examining the influence of the diffuse galaxy background.

We further required those clumps and catalogue objects to have measured fluxes above the average $5\sigma$ point-source depth limits of the filter bands. This reduced the number of detected simulated clumps from 7,161 to 3,385 and the number of point-like catalogue objects from 611 to 150.

\begin{figure}
    \centering
    \subfloat[\centering Star-like catalogue objects. \label{fig:hsc_phot_comp_diff_aper_stars}]{{\includegraphics[width=0.5\columnwidth]{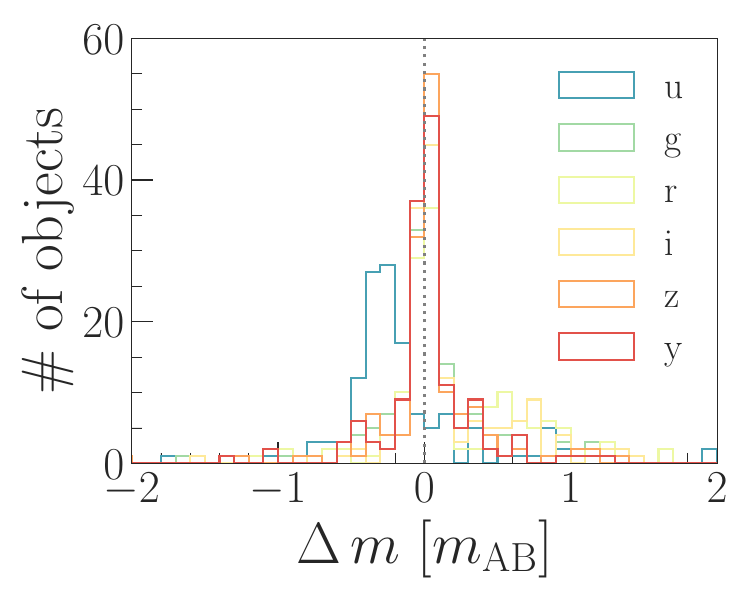} }}
    \subfloat[\centering Simulated clumps. \label{fig:hsc_phot_comp_diff_aper}]{{\includegraphics[width=0.5\columnwidth]{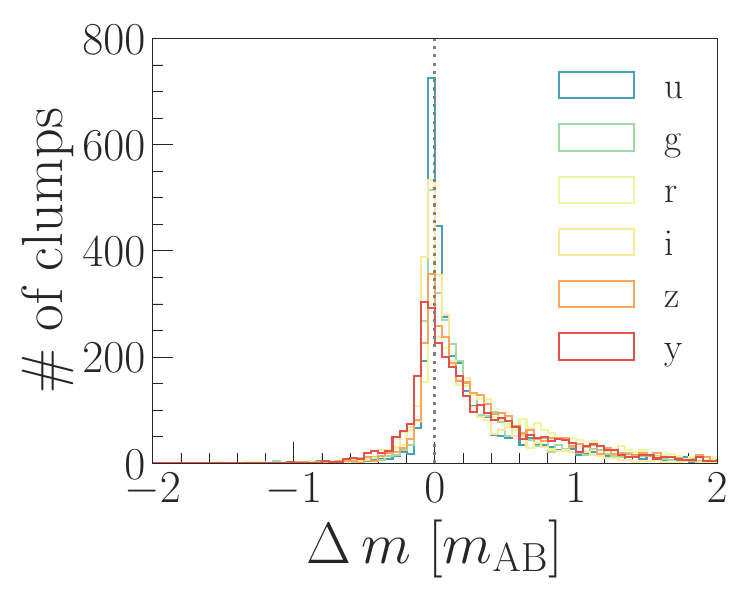} }}
    \caption[Histograms of the differences between recovered and true magnitudes of point-like catalogue objects and simulated clumps.]{Histograms of the differences between recovered and true magnitudes of point-like catalogue objects (a) and simulated clumps (b) that are brighter than the filter band-specific detection limits ($u\leq 27.1\,m_{\mathrm{AB}}$, $g\leq 26.5\,m_{\mathrm{AB}}$, $r\leq 26.1\,m_{\mathrm{AB}}$, $i\leq 25.9\,m_{\mathrm{AB}}$, $z\leq 25.1\,m_{\mathrm{AB}}$ and $y\leq 24.4\,m_{\mathrm{AB}}$).}
    \label{fig:hsc_phot_comp_diff_aper_star_sim}
\end{figure}

Using the final parameter values for the aperture radius, the annulus size and the radius of the clump mask, we measured the flux of all simulated clumps and the sample of star-like objects from the CLAUDS and HSC-SSP catalogues. The aperture was centred on the RA/DEC coordinates of the catalogue objects, but we used the positions of the identified flux peaks within each bounding box for the simulated clumps (see Section \ref{sec:clump_identification}). We do this, rather than using the known true position of the simulated clumps, because we also wanted to include the potential errors that were induced by our peak finding algorithm. In Figure \ref{fig:hsc_phot_comp_diff_aper_star_sim}, we show histograms of the differences ($\Delta m$) between the measured magnitudes and the true magnitudes of the point-like catalogue objects (Fig. \ref{fig:hsc_phot_comp_diff_aper_stars}) and simulated clumps (Fig. \ref{fig:hsc_phot_comp_diff_aper}).

For both test samples, the modes of the distributions are located around the value $\Delta m = 0.0$ in each filter band. The mode for the \textit{u}-band differences for the sample of star-like objects is slightly shifted to negative $\Delta m$ values, which indicates that the recovered \textit{u}-band magnitudes are underestimated. This is most likely due to the estimate of the ePSF for the CLAUDS images, which does not perfectly match the PSF used for the original CLAUDS catalogue values. The distributions of the magnitude differences from the sample of simulated clumps show an extended tail towards positive $\Delta m$ values (Fig. \ref{fig:hsc_phot_comp_diff_aper}), indicating that the recovered magnitudes tend to be overestimated most likely due to an insufficient background subtraction. We observed those larger differences mainly for simulated clumps that are close to the detection limit. In general, the measured values of the aperture photometry were found to be accurate within the $2\sigma$ error margins in 77\% to 81\% cases for all filter bands.

We note, however, that the sizes of the apertures we used to measure the fluxes of our detected clumps can exceed the physical sizes reported for clumps from high-redshift galaxies \citep[$\sim 1$ kpc, e.g.][]{Elmegreen2007a,FoersterSchreiber2011a,Guo2018,Zanella2019}. With the given seeing, features with physical sizes $<1$ kpc can be theoretically resolved in only a small fraction of our galaxy sample at $z\lesssim 0.1$. Due to the limited spatial resolution, the vast majority of the clumps are unresolved in our observations and might consist of blended smaller objects or still include some contamination from intra-clump areas of the galactic disk.

\section{The fraction of clumpy galaxies}\label{sec:clumpy_fraction}
To be able to directly compare the clumpy galaxies and the calculated clumpy fraction with the results from \citet{Guo2018} and \citet{Adams2022}, we limited our galaxy sample to those galaxies that have data from all six \textit{ugrizy}-filter bands, a $\mathrm{sSFR} \geq 10^{-11}\,\mathrm{yr}^{-1}$ and excluded all (close to) edge-on galaxies that have an elongation of $>3.0$, which we determined from our remeasurement of the morphological galaxy parameters (Appendix \ref{sec:data_preprocess_reff}). We note, however, that our remeasurement of the morphological galaxy parameters is based on a flux threshold and some SFGs could have been wrongly identified either as (close to) edge-on or with an elongation of $<3.0$ if extended low-surface brightness areas of the target galaxies were not included in our segmentation mask. However, given our mass-completeness limit of $\log(M_\star/M_\odot) \geq 9.0$, we expect only a minor influence of this error on our results. To estimate the fraction of galaxies with possible wrong morphological parameters, we visually inspected the 368 SFGs that were determined as (close to) edge-on and excluded from our sample. We find that five galaxies (or 1.4\%) appear to be less inclined and were wrongly excluded. A similar vetting of a random sample of 960 galaxies that are included in our sample results in a smaller fraction of galaxies with wrong morphological parameters (seven galaxies or 0.7\%).

The galaxies of our sample were further grouped into stellar mass bins that span the same mass ranges used by \citet{Guo2018} and \citet{Adams2022}. The three mass bins are defined for galaxy stellar masses (1) between $9.0 \leq \log(M_\star/M_\odot) < 9.8$, (2) between $9.8 \leq \log(M_\star/M_\odot) < 10.6$ and (3) between $10.6 \leq \log(M_\star/M_\odot) < 11.4$. Bins (2) and (3) are mass-complete over the whole observed redshift range, whereas bin (1) is mass-complete for redshift $z<0.32$. The mass bin boundaries are also plotted in Figure \ref{fig:hsc_data_galaxy_completeness} and the counts of galaxies, SFGs, clumpy SFGs and clumps are shown in Table \ref{tab:hsc_phys_props_galaxy_clumpy_fraction_counts}. For each mass bin we list total counts as well as counts for clumps that have a clump-galaxy flux ratio $F_{u,\mathrm{cl}}/F_{u,\mathrm{gal}}$ of $\geq0.08$.

We calculated the \textit{u}-band flux ratio $F_{u,\mathrm{cl}}/F_{u,\mathrm{gal}}$ as the ratio of the measured clump \textit{u}-band flux $F_{u,\mathrm{cl}}$ divided by the \textit{u}-band flux of the host galaxy $F_{u,\mathrm{gal}}$, which we obtained from the CLAUDS Source Extractor photometry catalogue \citep{Desprez2023,Picouet2023}. To account for the uncertainty of $F_{u,\mathrm{cl}}/F_{u,\mathrm{gal}}$, we used a Monte Carlo method to include contributions from the error estimates of our aperture photometry measurements and the published errors of the \textit{u}-band fluxes of the host galaxies. For each clump, we drew 100 random trials from normal distributions defined by the standard error values of $F_{u,\mathrm{cl}}$ and $F_{u,\mathrm{gal}}$ and determined the clumpy fraction using Equation \ref{eq:gsfc_fclumpy} for the different redshift and stellar mass bins of our sample of SFGs. Here, we only considered off-centre clumps that are further than 2\% of the Petrosian radius of the host galaxy ($d \geq 0.02\,R_P$) from the galactic centre to count the number of clumpy SFGs.

\begin{table}
	\centering
	\caption[Number of galaxies, SFGs and clumpy SFGs and clumps per host galaxy mass bin.]{Number of galaxies, SFGs ($\mathrm{sSFR} \geq 10^{-11}\,\mathrm{yr}^{-1}$), clumpy SFGs and clumps per host galaxy mass bin ($z\leq0.32$). The three mass bins are defined for galaxy stellar masses between $9.0 \leq \log(M_\star/M_\odot) < 9.8$ (low), between $9.8 \leq \log(M_\star/M_\odot) < 10.6$ (medium) and between $10.6 \leq \log(M_\star/M_\odot) < 11.4$ (high).}
    \label{tab:hsc_phys_props_galaxy_clumpy_fraction_counts}
	\footnotesize
        \begin{tabular}{lrrrrr}
		\hline
        Mass bin & \multicolumn{1}{l}{Galaxies} & \multicolumn{1}{l}{SFGs} & \multicolumn{1}{l}{SFGs} & \multicolumn{1}{l}{Clumps} & \multicolumn{1}{l}{Clumps/} \\
                  & \multicolumn{1}{l}{(all)} & & \multicolumn{1}{l}{(clumpy)} &                            & \multicolumn{1}{l}{galaxy}  \\
		\hline
        \multicolumn{6}{l}{with U-band flux ratio $F_{u,\mathrm{cl}}/F_{u,\mathrm{gal}}\geq0.08$} \\
        \hline
        Low    & 2,073 & 2,064 & 383 & 546 & 1.43 \\
        Medium & 3,002 & 2,528 & 382 & 580 & 1.52 \\
        High   & 2,703 &   803 & 101 & 153 & 1.51 \\
        All    & 7,778 & 5,395 & 866 & 1,279 & 1.48 \\
		\hline
        \multicolumn{6}{l}{Total} \\
        \hline
        Low    & 2,073 & 2,064 & 1,256 & 3,797 & 3.02 \\
        Medium & 3,002 & 2,528 & 1,695 & 6,798 & 4.01 \\
        High   & 2,703 &   803 & 513 & 2,195 & 4.28 \\
        All    & 7,778 & 5,395 & 3,464 & 12,790 & 3.69 \\
        \hline
	\end{tabular}
\end{table}

\subsection{Completeness of the clump detections}\label{sec:clumpy_fraction_incomplete}
\begin{figure}
    \centering
    \subfloat[\centering \textit{u}-band apparent magnitude. \label{fig:hsc_phys_props_galaxy_completeness_clumps_a}]{{\includegraphics[width=0.5\columnwidth]{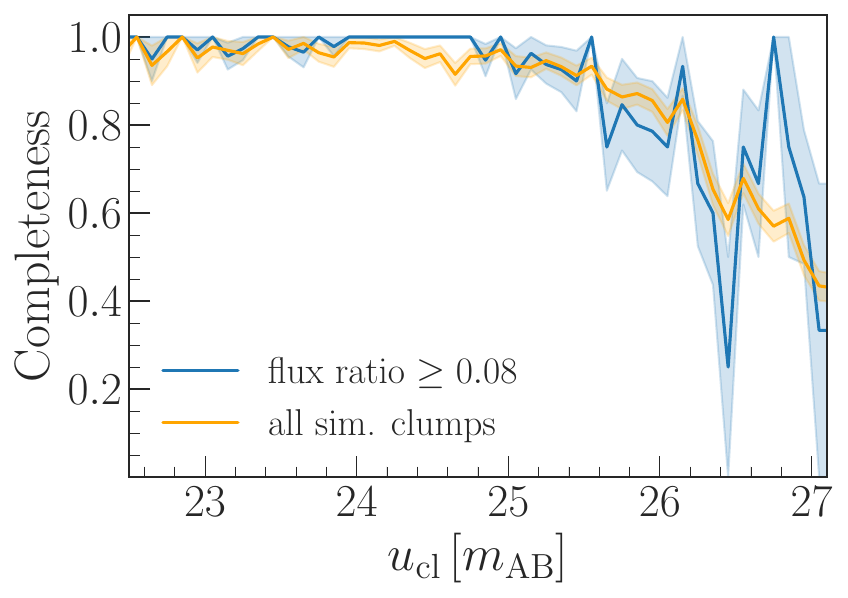} }}
    \subfloat[\centering Colour (\textit{u}-\textit{r}). \label{fig:hsc_phys_props_galaxy_completeness_clumps_b}]{{\includegraphics[width=0.5\columnwidth]{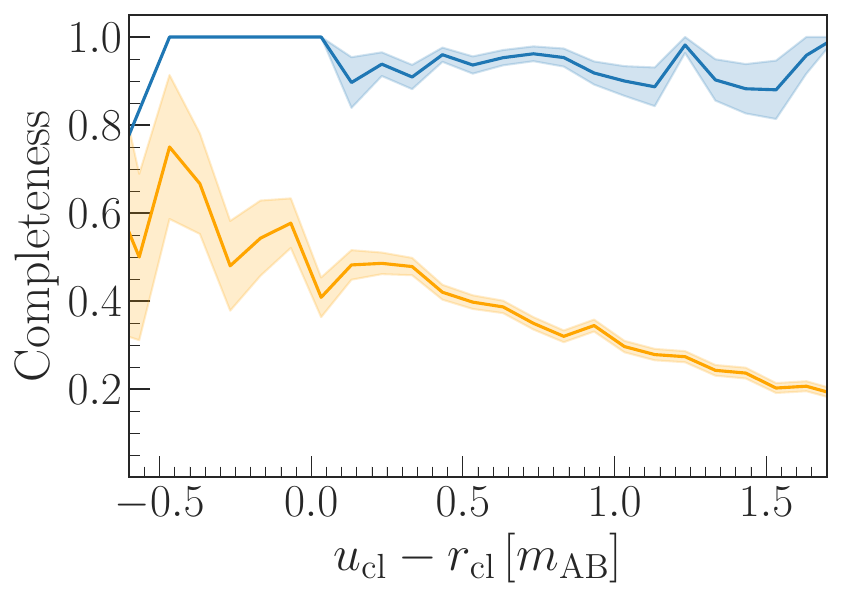} }}
    \\
    \subfloat[\centering Colour contrast \textit{u}-band. \label{fig:hsc_phys_props_galaxy_completeness_clumps_c}]{{\includegraphics[width=0.5\columnwidth]{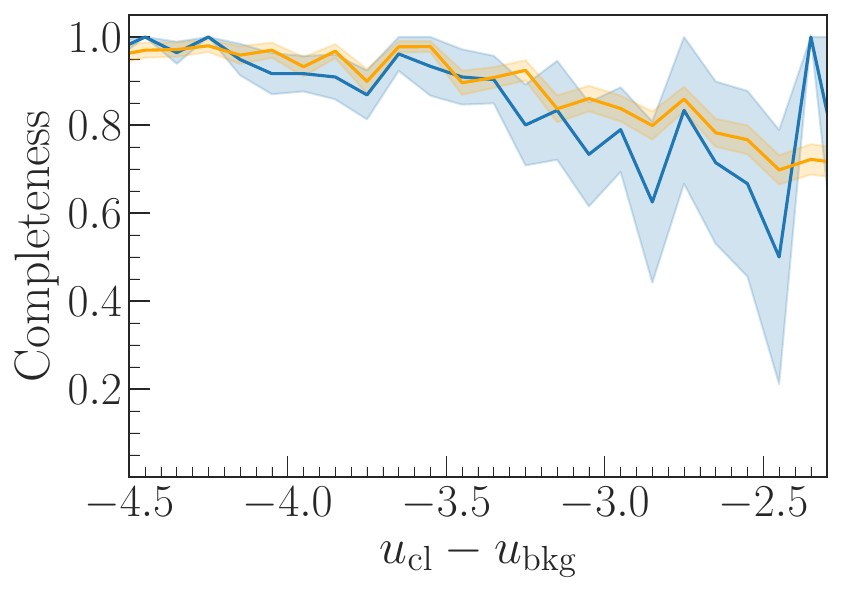} }}
    \subfloat[\centering Radial distance. \label{fig:hsc_phys_props_galaxy_completeness_clumps_d}]{{\includegraphics[width=0.5\columnwidth]{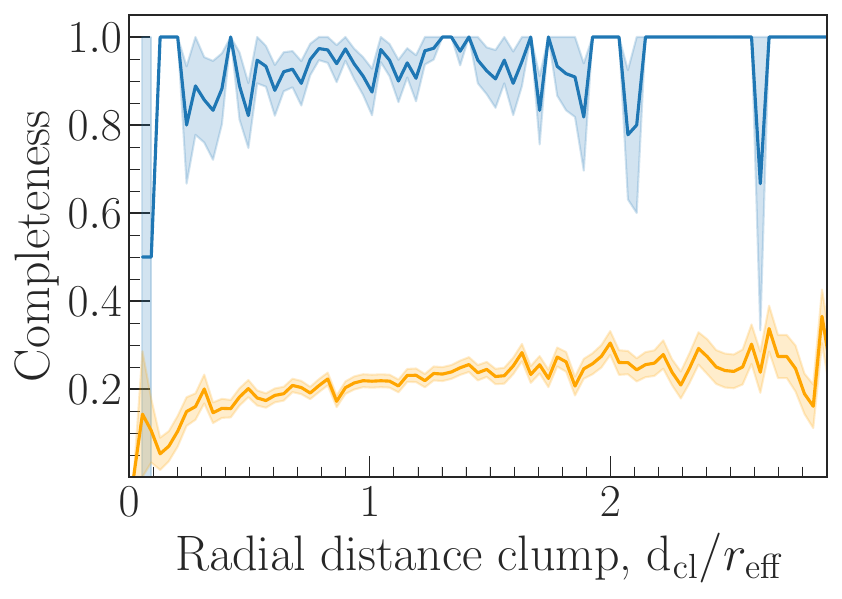} }}
    \\
    \subfloat[\centering Simulated clump stellar mass. \label{fig:hsc_phys_props_galaxy_completeness_clumps_e}]{{\includegraphics[width=0.5\columnwidth]{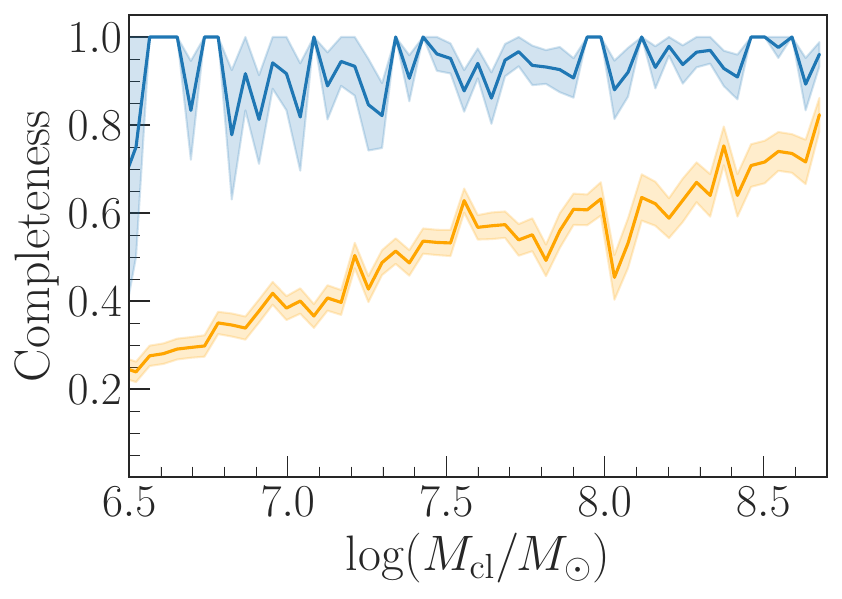} }}
    \subfloat[\centering Mass ratio clump/galaxy. \label{fig:hsc_phys_props_galaxy_completeness_clumps_f}]{{\includegraphics[width=0.5\columnwidth]{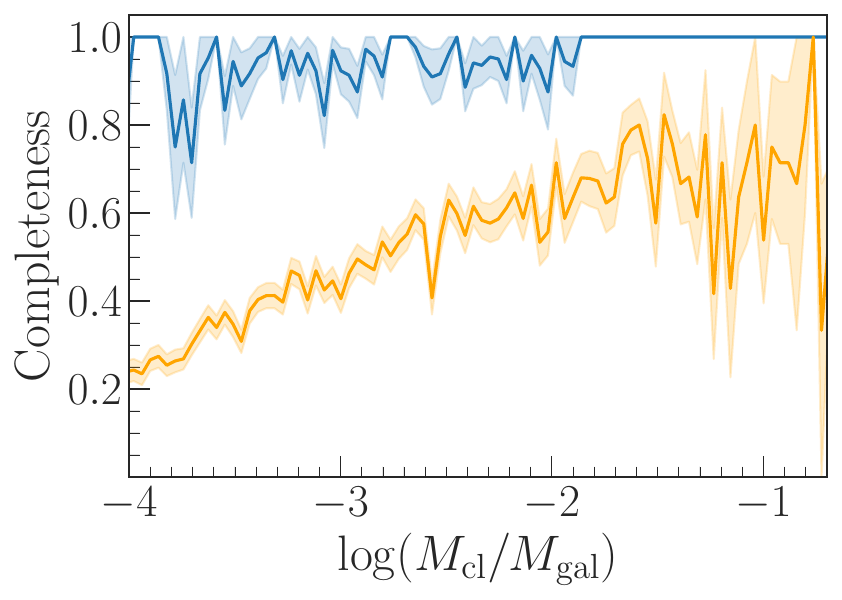} }}
    \caption[Detection completeness of the simulated clumps as a function of different physical clump parameters and per clump-galaxy ratio $F_{u,\mathrm{cl}}/F_{u,\mathrm{gal}} \geq 0.08$.]{Detection completeness of the simulated clumps as a function of different physical clump parameters. The completeness is plotted in orange for the full sample of simulated clumps and in blue for simulated clumps with a clump-galaxy \textit{u}-band flux ratio of $\geq 0.08$. The shaded areas show the $1\sigma$ errors.}
    \label{fig:hsc_phys_props_galaxy_completeness_clumps}
\end{figure}

The completeness of the object detection model was validated using simulated clumps that were injected into real galaxy images \citep[see Appendix \ref{sec:frcnn_development} and][]{Popp2026b}. The overall completeness is relatively low (22.21\%) because the sample of simulated clumps included many faint objects. However, after applying the clump-galaxy \textit{u}-band flux ratio thresholds $F_{u,\mathrm{cl}}/F_{u,\mathrm{gal}} \geq 0.08$, the detection completeness is increased. Figure \ref{fig:hsc_phys_props_galaxy_completeness_clumps} plots the completeness as a function of the \textit{u}-band flux (Fig. \ref{fig:hsc_phys_props_galaxy_completeness_clumps_a}), the colour (\textit{u}-\textit{r}) (Fig. \ref{fig:hsc_phys_props_galaxy_completeness_clumps_b}), the magnitude difference between the clump's flux and the estimated underlying background flux in the \textit{u}-band (Fig. \ref{fig:hsc_phys_props_galaxy_completeness_clumps_c}, measured from an annulus around the clump's position, see also Section \ref{sec:clump_phot}), the relative radial distance from the galaxy centre (Fig. \ref{fig:hsc_phys_props_galaxy_completeness_clumps_d}), the stellar mass of the clump (Fig. \ref{fig:hsc_phys_props_galaxy_completeness_clumps_e}) and the ratio of the clump's stellar mass to the host galaxy's mass (Fig. \ref{fig:hsc_phys_props_galaxy_completeness_clumps_f}) for all simulated clumps as well as for those above the flux ratio threshold. The detection completeness for clumps that are above the threshold is $>0.8$ over most of the simulated ranges of the different parameters.

The selection of bright clumps using a clump-galaxy flux ratio thresholds clearly improves the detection completeness, irrespective of colour (\textit{u}-\textit{r}), stellar mass or the clump-galaxy mass ratio of a clump. Detections are also improved if the contrast between the clump and the underlying galactic background is high (parametrised as the \textit{u}-band magnitude difference in Figure \ref{fig:hsc_phys_props_galaxy_completeness_clumps_c}). Close to the galactic centre, the completeness of the clump detections starts to drop significantly (Fig. \ref{fig:hsc_phys_props_galaxy_completeness_clumps_d}). Therefore, clumps that are less than $\mathrm{d}_{\mathrm{cl}} < 0.3\,r_{\mathrm{eff}}$ from the galaxy centre are excluded from our analysis. Here, the distance $\mathrm{d}_{\mathrm{cl}}$ is measured in units of effective or half-light radius $r_{\mathrm{eff}}$ of the galaxy (Appendix \ref{sec:data_preprocess_reff}) from the flux-weighted centroid of the source object that was determined by the HSC pipeline and its coordinates are available from the HSC-SSP PDR3 catalogue \citep{Bosch2017}.

\begin{figure}
    \centering
    \subfloat[\centering Host galaxy redshift. \label{fig:hsc_phys_props_galaxy_completeness_gal_a}]{{\includegraphics[width=0.5\columnwidth]{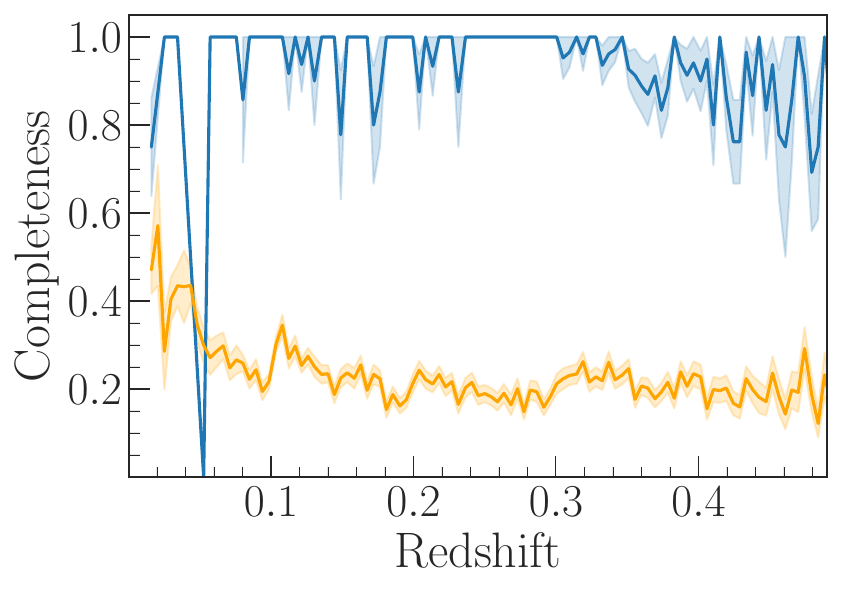} }}
    \subfloat[\centering Host galaxy stellar mass. \label{fig:hsc_phys_props_galaxy_completeness_gal_b}]{{\includegraphics[width=0.5\columnwidth]{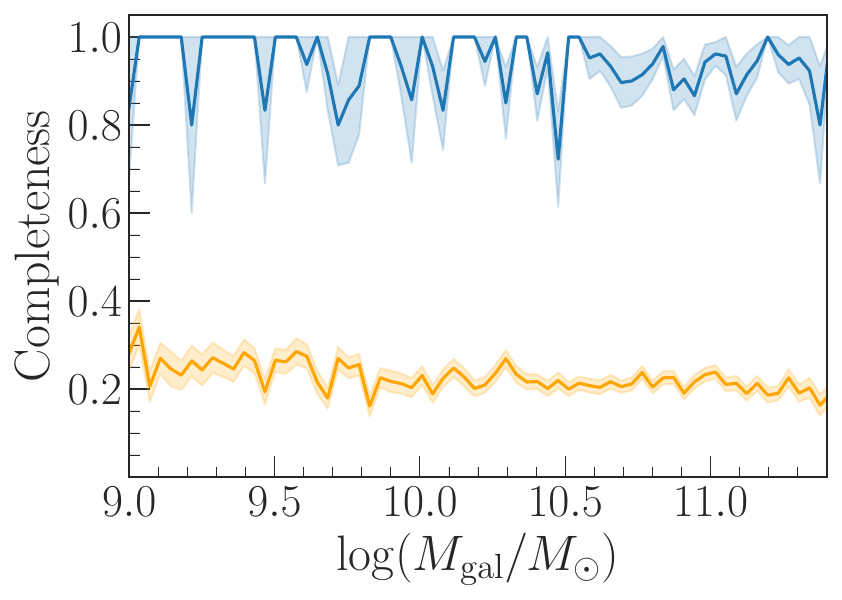} }}
    \\
    \subfloat[\centering Host galaxy sSFR. \label{fig:hsc_phys_props_galaxy_completeness_gal_c}]
    {{\includegraphics[width=0.5\columnwidth]{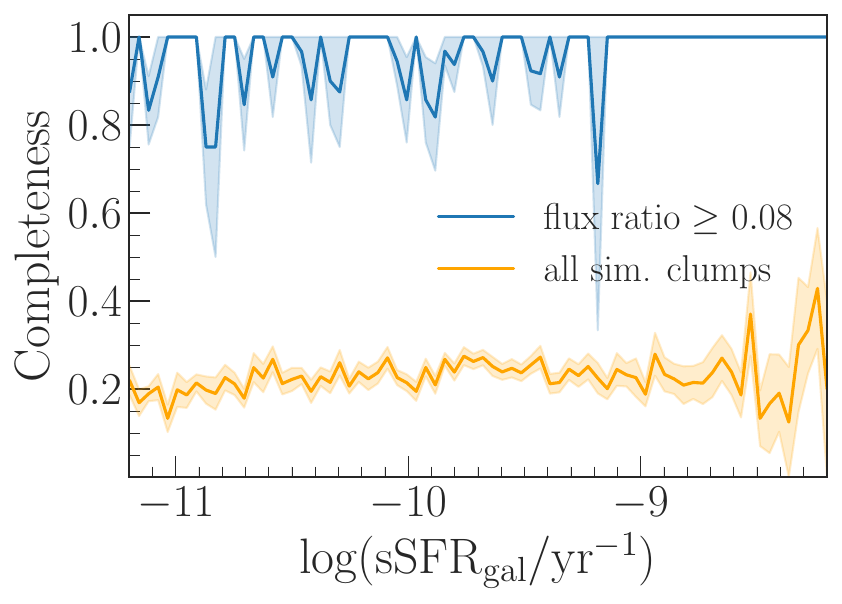} }}
    \caption[Detection completeness of the simulated clumps as a function of different host galaxy parameters and per clump-galaxy ratio $F_{u,\mathrm{cl}}/F_{u,\mathrm{gal}} \geq 0.08$.]{Similar to Figure \ref{fig:hsc_phys_props_galaxy_completeness_clumps} but showing the detection completeness of the simulated clumps as a function of different host galaxy parameters.}
    \label{fig:hsc_phys_props_galaxy_completeness_gal}
\end{figure}

The detection completeness versus the physical properties of the host galaxies, for clumps brighter than the clump-galaxy flux ratio threshold, is also $>0.8$ and much higher than for the full sample of simulated clumps (Figure \ref{fig:hsc_phys_props_galaxy_completeness_gal}). In particular, the completeness is high for the host galaxy stellar mass ranges (Fig. \ref{fig:hsc_phys_props_galaxy_completeness_gal_b}) that define the stellar mass bins used for the comparison of our clump sample to the samples from \citet{Guo2018} and \citet{Adams2022}.

Even though the detection completeness for clumps with a clump-galaxy \textit{u}-band flux ratio of $F_{u,\mathrm{cl}}/F_{u,\mathrm{gal}} \geq 0.08$ is high, it is not unity. Therefore, the sample of detected clumpy galaxies is likely to be incomplete due to some undetected bright clumps. To correct for the incompleteness, we followed a similar approach to that described in \citet{Adams2022}. The details of our approach are described in Appendix \ref{sec:clumpy_fraction_incomplete_calc}. The incompleteness correction increased our observed clumpy fractions by $\lesssim0.3\%$ across all redshift and stellar mass bins (Table \ref{tab:hsc_phys_props_galaxy_clumpy_fraction_low_z}) indicating that our detection method had failed to detect any clumps in only a very small fraction of clumpy SFGs.

\subsection{Observed clumpy fraction}\label{sec:clumpy_fraction_obs}
\begin{figure}
    \centering
    \includegraphics[width=1\columnwidth]{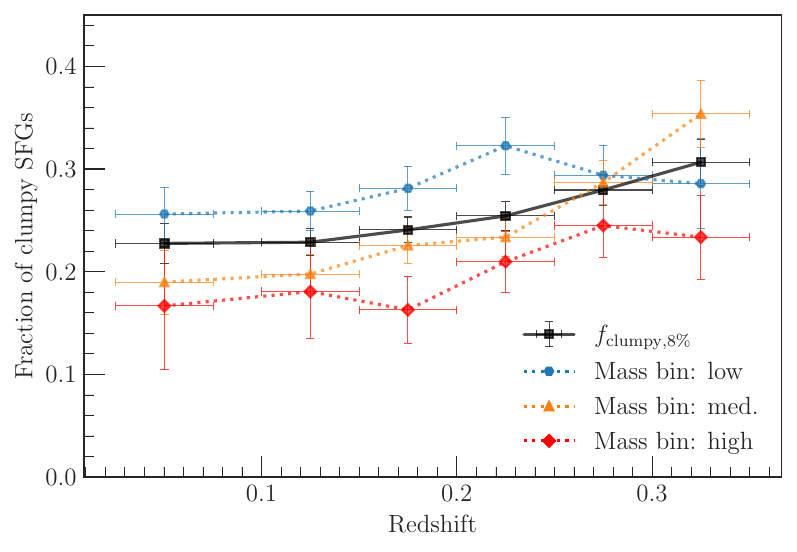}
    \caption[Incompleteness-corrected clumpy fraction as a function of redshift for SFGs in different stellar mass bins and clumps defined by a clump-galaxy ratio $F_{u,\mathrm{cl}}/F_{u,\mathrm{gal}} \geq 0.08$.]{Incompleteness-corrected clumpy fraction as a function of redshift for SFGs in different stellar mass bins and clumps defined by a clump-galaxy flux ratio $F_{u,\mathrm{cl}}/F_{u,\mathrm{gal}} \geq 0.08$ ($f_{\mathrm{clumpy,8\%}}$). The low-mass galaxies ($9.0 \leq \log(M_\star/M_\odot) < 9.8$) are shown in blue, the medium-mass galaxies ($9.8 \leq \log(M_\star/M_\odot) < 10.6$) in orange and the high-mass galaxies ($10.6 \leq \log(M_\star/M_\odot) < 11.4$) in red. The clumpy fraction over all galaxy stellar mass bins are shown in black. Error bars show the standard error for the clumpy fractions and the bin range for redshift.}
    \label{fig:hsc_phys_props_galaxy_clumpy_fraction_low_z}
\end{figure}

In Figure \ref{fig:hsc_phys_props_galaxy_clumpy_fraction_low_z} and Table \ref{tab:hsc_phys_props_galaxy_clumpy_fraction_low_z}, we show the observed and incompleteness-corrected clumpy fractions from 5,395 SFGs with redshift $z\leq 0.32$ from the CLAUDS/HSC mass-complete sample (of which 3,464 contain at least one off-centre clump), together with their corresponding binomial errors. The clumpy fractions are plotted for each of the mass bins, together with the aggregated clumpy fraction for clumps with $F_{u,\mathrm{cl}}/F_{u,\mathrm{gal}} \geq 0.08$ and galaxy stellar masses of $M_\star \geq 10^9\,M_\odot$. The redshift range is limited to $z \leq 0.32$, so that the galaxy sample is mass-complete for all bins.

\begin{table}
	\centering
	\caption[Observed and incompleteness-corrected values of the clumpy fraction.]{Observed and incompleteness-corrected values of the clumpy fractions. The values are listed in percent for each of the mass-complete bins and redshift range of the SFGs sample ($\mathrm{sSFR} \geq 10^{-11}\,\mathrm{yr}^{-1}$). The three mass bins are defined for galaxy stellar masses between $9.0 \leq \log(M_\star/M_\odot) < 9.8$ (low), between $9.8 \leq \log(M_\star/M_\odot) < 10.6$ (medium) and between $10.6 \leq \log(M_\star/M_\odot) < 11.4$ (high).}
    \label{tab:hsc_phys_props_galaxy_clumpy_fraction_low_z}
	\footnotesize
        \begin{tabular}{lrrr}
		\hline
        Mass bin & \multicolumn{1}{c}{Redshift} & \multicolumn{2}{c}{$f_{\mathrm{clumpy,8\%}}$} \\
                 & \multicolumn{1}{c}{$z$} & \multicolumn{1}{c}{Observed} & \multicolumn{1}{c}{Corrected} \\
		\hline
        Low & $[0.0005, 0.1]$ & $25.26\pm 2.57\%$ & $25.61\pm 2.59\%$ \\
        Low & $(0.1, 0.15]$   & $25.71\pm 1.91\%$ & $25.90\pm 1.91\%$ \\
        Low & $(0.15, 0.2]$   & $27.89\pm 2.14\%$ & $28.12\pm 2.14\%$ \\
        Low & $(0.2, 0.25]$   & $31.90\pm 2.79\%$ & $32.26\pm 2.80\%$ \\
        Low & $(0.25, 0.3]$   & $28.98\pm 2.90\%$ & $29.39\pm 2.91\%$ \\
        Low & $(0.3, 0.32]$   & $28.57\pm 4.41\%$ & $28.57\pm 4.41\%$ \\
        \hline
        Medium & $[0.0005, 0.1]$ & $18.99\pm 3.12\%$ & $18.99\pm 3.12\%$ \\
        Medium & $(0.1, 0.15]$   & $19.54\pm 1.90\%$ & $19.77\pm 1.91\%$ \\
        Medium & $(0.15, 0.2]$   & $22.38\pm 1.73\%$ & $22.55\pm 1.73\%$ \\
        Medium & $(0.2, 0.25]$   & $23.18\pm 1.85\%$ & $23.37\pm 1.85\%$ \\
        Medium & $(0.25, 0.3]$   & $28.47\pm 2.15\%$ & $28.70\pm 2.16\%$ \\
        Medium & $(0.3, 0.32]$   & $34.91\pm 3.27\%$ & $35.38\pm 3.28\%$ \\
        \hline
        High & $[0.0005, 0.1]$ & $16.67\pm 6.21\%$ & $16.67\pm 6.21\%$ \\
        High & $(0.1, 0.15]$   & $18.06\pm 4.53\%$ & $18.06\pm 4.53\%$ \\
        High & $(0.15, 0.2]$   & $16.28\pm 3.25\%$ & $16.28\pm 3.25\%$ \\
        High & $(0.2, 0.25]$   & $20.97\pm 2.98\%$ & $20.97\pm 2.98\%$ \\
        High & $(0.25, 0.3]$   & $24.49\pm 3.07\%$ & $24.49\pm 3.07\%$ \\
        High & $(0.3, 0.32]$   & $23.36\pm 4.09\%$ & $23.36\pm 4.09\%$ \\
        \hline
        All & $[0.0005, 0.1]$ & $22.55\pm 1.91\%$ & $22.76\pm 1.92\%$ \\ 
        All & $(0.1, 0.15]$   & $22.58\pm 1.30\%$ & $22.87\pm 1.31\%$ \\ 
        All & $(0.15, 0.2]$   & $23.81\pm 1.26\%$ & $24.07\pm 1.26\%$ \\ 
        All & $(0.2, 0.25]$   & $25.23\pm 1.38\%$ & $25.43\pm 1.39\%$ \\ 
        All & $(0.25, 0.3]$   & $27.73\pm 1.51\%$ & $27.95\pm 1.51\%$ \\ 
        All & $(0.3, 0.32]$   & $30.42\pm 2.23\%$ & $30.66\pm 2.24\%$ \\ 
        \hline
	\end{tabular}
\end{table}

The clumpy fraction that is aggregated over all mass bins increases slowly with increasing redshift from $f_{\mathrm{clumpy,8\%}} \sim 23\%$ at $z\sim 0.1$ to $\sim$31\% at $z\sim 0.3$. In comparison, the fraction of low-mass galaxies having at least one clump is $\sim$3\% higher and increases more strongly for redshift $>0.15$ before dropping below the aggregated clumpy fraction for SFGs in our highest redshift bin. The clumpy fraction for SFGs in the medium mass bin is $\sim$2\% lower than the aggregated clumpy fraction but we observe a stronger increase with increasing redshift for $z>0.2$. The $f_{\mathrm{clumpy,8\%}}$ values for the high-mass galaxies are $\sim$5\% lower than the aggregated clumpy fractions and also follow the generally increasing trend with redshift with a slight drop to $f_{\mathrm{clumpy,8\%}} \sim 23\%$ at $z\sim 0.3$.

\subsection{Clump catalogue}\label{sec:clump_cat}
We release a catalogue containing the detected clump candidates and their measured and derived properties along with this paper. The catalogue contains the clumps' coordinates together with measured fluxes and magnitudes for each of the \textit{ugrizy}-filter bands measured from the CLAUDS and HSC-SSP science images. The catalogue also contains estimates for the stellar mass, age, metallicity and dust attenuation of the clumps which we inferred through SED fitting. The SED fitting process and analysis of the physical clumps properties are described in detail in Popp et al. (2026c, submitted). Table \ref{tab:catalogue} describes the released catalogue in compact form.
\begin{table*}
	\centering
	\caption[Description of the clump catalogue for the CLAUDS/HSC-SSP galaxies.]{Description of the clump catalogue for the CLAUDS/HSC-SSP galaxies.} 
    \label{tab:catalogue}
	\begin{tabular}{llll}
		\hline
		Field & Description & Units & Source \\
		\hline
        HSCobjid & HSC-SSP PDR3 Object-ID & & HSC-SSP PDR3 \\
        bbox\_id & Bounding box-ID & & \\
        clump\_id & Clump-ID & & \\
        objectness & Objectness score  & & \\
        clump\_ra & Right Ascension & Degrees & \\
        clump\_dec & Declination & Degrees & \\
        clump\_aper\_corr\_(u|g|r|i|z|y) & Aperture correction factor & & \\
        clump\_flux\_nJy\_(u|g|r|i|z|y) & Measured flux & nJy & \\
        clump\_flux\_nJy\_err\_(u|g|r|i|z|y) & Error measured flux & nJy & \\
        clump\_flux\_nJy\_bkgsub\_(u|g|r|i|z|y) & Bkg.-subtracted flux & nJy & \\
        clump\_flux\_nJy\_err\_bkgsub\_(u|g|r|i|z|y) & Error bkg.-subtracted flux & nJy & \\
        clump\_flux\_ABmag\_(u|g|r|i|z|y) & Clump magnitude (measured) & AB mag & \\
        clump\_flux\_ABmag\_err\_(u|g|r|i|z|y) & Error clump magnitude (measured) & $m_{\mathrm{AB}}$ & \\  
        clump\_flux\_ABmag\_bkgsub\_(u|g|r|i|z|y) & Clump magnitude (bkg.-subtracted) & $m_{\mathrm{AB}}$ & \\
        clump\_flux\_ABmag\_err\_bkgsub\_(u|g|r|i|z|y) & Error clump magnitude (bkg.-subtracted) & $m_{\mathrm{AB}}$ & \\
        clump\_flux\_ABmag\_bkgsub\_corr\_(u|g|r|i|z|y) & Clump mag. (bkg.-subtracted and ext. corr.) & $m_{\mathrm{AB}}$ & \\
        clump\_norm\_distance\_r\_eff & Normalised galactocentric distance & $r_{\mathrm{eff}}$ & \\
        clump\_mass\_(MAP|median|mode) & Point estimate clump stellar mass & $M_\odot$ & \\
        clump\_mass\_(MAP|median|mode)\_lower & Lower bound 95\% credible region stellar mass & $M_\odot$ & \\
        clump\_mass\_(MAP|median|mode)\_upper & Upper bound 95\% credible region stellar mass & $M_\odot$ & \\
        clump\_logzsol\_(MAP|median|mode) & Point estimate clump metallicity & $\log(Z/Z_\odot)$ & \\
        clump\_logzsol\_(MAP|median|mode)\_lower & Lower bound 95\% credible region metallicity & $\log(Z/Z_\odot)$ & \\
        clump\_logzsol\_(MAP|median|mode)\_upper & Upper bound 95\% credible region metallicity & $\log(Z/Z_\odot)$ & \\
        clump\_dust2\_(MAP|median|mode) & Point estimate clump dust attenuation $A_{V}$ & $m_{\mathrm{AB}}$ & \\
        clump\_dust2\_(MAP|median|mode)\_lower & Lower bound 95\% credible region $A_{V}$ & $m_{\mathrm{AB}}$ & \\
        clump\_dust2\_(MAP|median|mode)\_upper & Upper bound 95\% credible region $A_{V}$ & $m_{\mathrm{AB}}$ & \\
        clump\_tage\_(MAP|median|mode) & Point estimate clump stellar age & Gyr & \\
        clump\_tage\_(MAP|median|mode)\_lower & Lower bound 95\% credible region stellar age & Gyr & \\
        clump\_tage\_(MAP|median|mode)\_upper & Upper bound 95\% credible region stellar age & Gyr & \\
        clump\_SFR\_MAP & Point estimate clump star-formation rate & $M_\odot\,\mathrm{yr}^{-1}$ & \\
        clump\_sSFR\_MAP & Point estimate clump specific star-formation rate & $\mathrm{yr}^{-1}$ & \\
        clump\_mfrac\_MAP & Fraction of survived clump stellar mass & & \\
        gal\_cmodel\_flux\_nJy\_(u|g|r|i|z|y) & Galaxy flux & nJy & CLAUDS/HSC-SSP PDR3 \\
        gal\_cmodel\_flux\_ABmag\_(u|g|r|i|z|y) & Galaxy magnitude & $m_{\mathrm{AB}}$ & CLAUDS/HSC-SSP PDR3 \\
        gal\_mass & Galaxy stellar mass & $\log(M_\star/ M_\odot)$ & CLAUDS/HSC-SSP PDR3 \\
        gal\_sSFR & Galaxy specific star-formation rate & $\log(\mathrm{yr}^{-1})$ & CLAUDS/HSC-SSP PDR3 \\
        gal\_source\_elong & Galaxy elongation & & \\
        gal\_redshift & Galaxy redshift & & SDSS, CLAUDS/HSC-SSP PDR3 \\
        gal\_redshift\_error & Galaxy redshift error & & SDSS, CLAUDS/HSC-SSP PDR3 \\l\_is\_specz & Flag for spectroscopic redshift & & HSC-SSP PDR3 \\
		\hline
	\end{tabular}
\end{table*}

When using the catalogue, we recommend applying the selection criteria for a mass-complete galaxy sample (see Section \ref{sec:data_gal_sample_compl}), including only SFGs (\texttt{gal\_sSFR} $\geq -11.0$) and excluding all galaxies with an elongation $>3.0$ (\texttt{gal\_source\_elong} $\leq 0.3$). Furthermore, all clumps that are too close to the galaxy centre (\texttt{clump\_norm\_distance\_r\_eff} $\geq 0.3$) are excluded from the analysis in this paper (Section \ref{sec:clumpy_fraction_incomplete}). Purity and completeness can be varied by filtering on the objectness or detection score (\texttt{objectness}), where a higher score threshold results in higher purity and vice versa \citep[see also][]{Popp2026b}. The objectness is a measure between 0.0 and 1.0 and indicates how certain the model is whether the detected object is a possible clump or not. 

\section{Comparison with other studies}\label{sec:discussion}
The main results of this paper are presented in Figure \ref{fig:hsc_phys_props_galaxy_clumpy_fraction_low_z} and Table \ref{tab:hsc_phys_props_galaxy_clumpy_fraction_low_z} for SFGs from different redshift and stellar mass bins. We determined the clumpy fraction in each galaxy bin using a similar definition of a massive star-forming clump adopted by \citet{Guo2015} for our sample of low-redshift SFGs. However, this definition is based on observations of high-redshift clumps located in SFGs around the peak of the cosmic SFR density. The observed low-redshift clumps from this study are not necessarily analogues of the high-redshift clumps and our observed clumpy fractions may not be consistent with the redshift evolution of $f_{\mathrm{clumpy}}$ expected from high-z SFGs. In the following sections we discuss our results in comparison to recent studies of clumpy galaxies from the literature.

\subsection{Properties of clumpy galaxies}\label{sec:clumpy_fraction_gals}
\begin{figure}
    \centering
    \includegraphics[width=1\columnwidth]{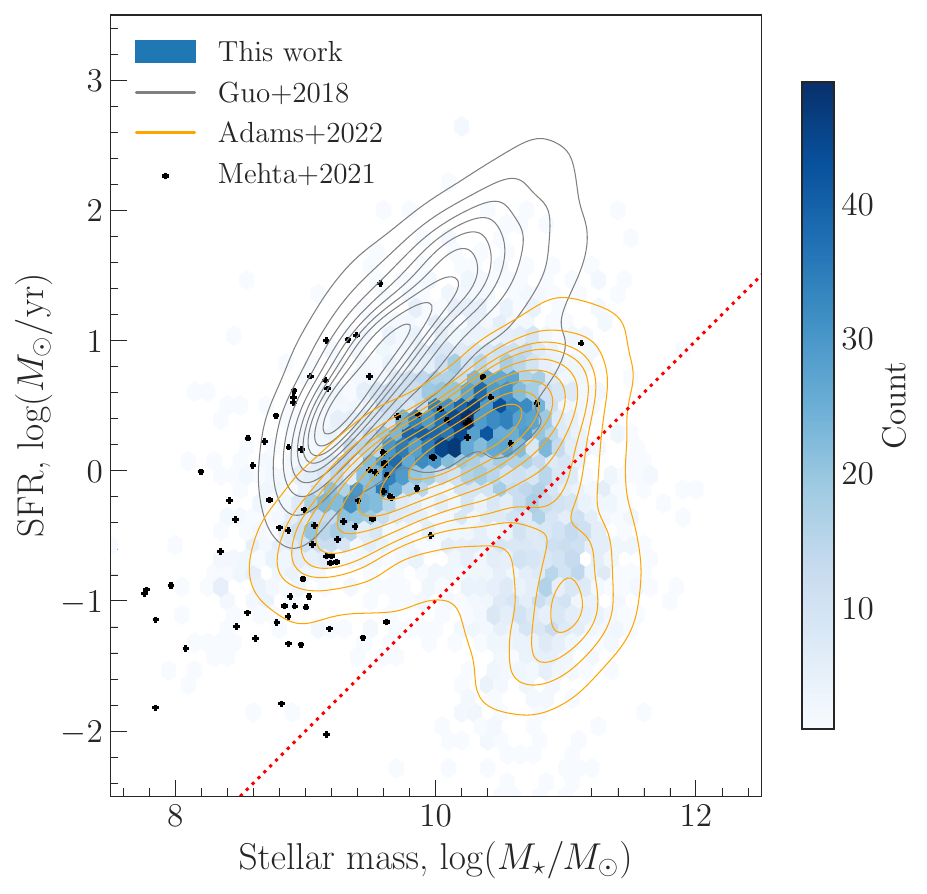}
    \caption[Galaxy SFR as a function of stellar mass for the CLAUDS and HSC-SSP galaxies with at least one off-centre clump.]{Galaxy SFR as a function of stellar mass for the CLAUDS and HSC-SSP galaxies with at least one off-centre clump. Also shown are the distributions of galaxies used in the studies from \citet{Guo2018,Mehta2021,Adams2022} for comparison. The distribution of clumpy galaxies from this study are shown in blue hex-bins. The galaxies analysed by \citet[][$0.5 \leq z \leq 3.0$]{Guo2018} are shown as grey contours, the galaxies from \citet[][$0.02 < z < 0.15$]{Adams2022} are plotted as orange contours and small black crosses mark the galaxy sample used by \citet[][$z < 0.06$]{Mehta2021}. The red dotted line marks the separation between star-forming and quiescent galaxies at $\mathrm{sSFR}=10^{-11}\,\mathrm{yr}^{-1}$.}
    \label{fig:hsc_phys_props_galaxy_pop_comparison}
\end{figure}

We first compared our sample of clumpy SFGs to the galaxy samples studied by \citet{Guo2018,Mehta2021,Adams2022}. We note that we also detected clumps in quiescent galaxies (i.e. $\mathrm{sSFR} < 10^{-11}\,\mathrm{yr}^{-1}$), which are excluded from the sample of galaxies we used to determine the clumpy fraction and are shown in Table \ref{tab:hsc_phys_props_galaxy_clumpy_fraction_counts}. However, we also plot those excluded clumpy galaxies in Figure \ref{fig:hsc_phys_props_galaxy_pop_comparison}, where we show the comparison of all galaxies used in the above-mentioned studies in a SFR/$M_\star$-diagram.

The 1,250 galaxies analysed by \citet{Guo2018} were selected from the CANDELS/GOODS-S survey \citep{Grogin2011,Koekemoer2011} for $M_\star>10^9\,M_\odot$, $\mathrm{sSFR}>10^{-10}\,\mathrm{yr}^{-1}$ and redshift between $0.5 \leq z \leq 3.0$. The galaxy sample from \citet{Mehta2021} were selected from the SDSS DR7 coverage of Stripe 82 and consists of 92 clumpy galaxies at redshift $z<0.06$ that span a stellar mass range of $10^{7.45} \leq M_\star \leq 10^{11.12}\,M_\odot$ and SFR range of $10^{-2.02} \leq \mathrm{SFR} \leq 10^{1.44}\,M_\odot \,\mathrm{yr}^{-1}$. The galaxy selection for \textit{Galaxy Zoo: Clump Scout} from \citet{Adams2022} includes 7,050 SDSS galaxies with redshift between $0.02 < z < 0.15$, with stellar masses of $10^{7.26} \leq M_\star \leq 10^{12.38}\,M_\odot$ and with $\mathrm{sSFR}>10^{-10}\,\mathrm{yr}^{-1}$.

The clumpy galaxies from \citet{Adams2022} are very similar to the sample used in this paper in terms of SFR and stellar mass but include a smaller redshift range for the host galaxies. The redshift range of the sample from \citet{Mehta2021} is even lower and the analysed galaxies are generally less massive but with similar SFRs to our CLAUDS/HSC sample. The high-redshift galaxies from \citet{Guo2018} show a significantly higher SFR than the low-redshift galaxies from the other samples. This is not unexpected as many of the galaxies from that sample are observed during an epoch of intense star-formation at $z\sim 2$ \citep[`cosmic noon',][]{Madau2014,FoersterSchreiber2020}.

\subsection{Redshift evolution of the clumpy fraction}\label{sec:clumpy_fraction_redshift}
\begin{figure}
    \centering
    \includegraphics[width=1\columnwidth]{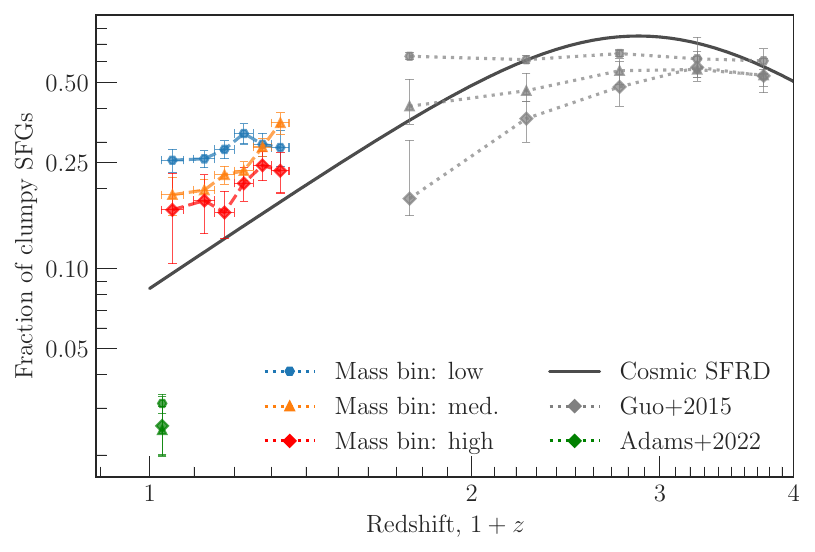}
    \caption[Incompleteness-corrected clumpy fraction as a function of redshift for SFGs in different stellar mass bins in comparison with \citet{Guo2015,Adams2022}.]{Incompleteness-corrected clumpy fraction as a function of redshift for SFGs in different stellar mass bins in comparison with \citet{Guo2015,Adams2022}. Similar to Fig. \ref{fig:hsc_phys_props_galaxy_clumpy_fraction_low_z} but with added clumpy fractions from \citet{Adams2022} in green and from \citet{Guo2015} in grey using the same symbols for the different stellar mass bins of the SFGs. Also shown is the evolution of the cosmic SFR density (SFRD) from \citet{Madau2014} in black but rescaled to arbitrary units for better comparison. Error bars show the standard error for the clumpy fractions and the bin range for redshift.}
    \label{fig:hsc_phys_props_galaxy_clumpy_fraction_mid_z}
\end{figure}

The redshift range we have probed with our galaxy sample is between the redshift ranges analysed by \citet{Guo2015} and \citet{Adams2022}. Figure \ref{fig:hsc_phys_props_galaxy_clumpy_fraction_mid_z} plots the clumpy fractions of SFGs that have at least one off-centre clump with $F_{u,\mathrm{cl}}/F_{u,\mathrm{gal}} \geq 0.08$ over the redshift range $0.0 \leq z \leq 3.0$ and compares our results to the results from \citet{Guo2015} and \citet{Adams2022}. The clumpy fractions are plotted separately for each of the three mass bins used by all authors. We note that the SDSS \textit{u}-band \citep[$\lambda_{\mathrm{eff}} = 360,8\,\mathrm{nm}$,][]{Gunn1998} and the CLAUDS \textit{u}-band \citep[$\lambda_{\mathrm{eff}} = 367.6\,\mathrm{nm}$ and $379.9\,\mathrm{nm}$ for the CFHT MegaCam filters $u$ and $u^\star$, respectively,][]{Sawicki2019} are not identical and \citet{Guo2015} define star-forming clumps based on the UV luminosity ratio. However, our comparison is motivated by tests from \citet{Adams2022} who showed that values for $f_{\mathrm{clumpy}}$ based on \textit{u}-band flux ratios are closely correlated with values for $f_{\mathrm{clumpy}}$ based on UV luminosity ratios.

Compared to the clumpy fractions from \citet{Adams2022}, we find a significantly increased clumpy fraction for the overlapping redshift range of $0.02 \leq z \leq 0.035$. We find $f_{\mathrm{clumpy}}\sim 23\%$ compared to the $\sim$3\% found by \citet{Adams2022}. This is likely related to the increased spatial resolution and sensitivity of the CLAUDS/HSC observations compared to the SDSS observations and the higher detection completeness achieved by our FRCNN model in comparison to models that were applied to SDSS data \citep{Popp2024}. Other studies also reported varying clumpy fractions for low-redshift galaxies. While a recent study by \citet{Chugunov2026} estimated $f_{\mathrm{clumpy}}\sim 22\%$ at redshift $z\sim0.2$ using HST observations, which aligns well with the results of this study, another study by \citet{Murata2014} reported a clumpy fraction of $\sim$5\% at $z \sim 0.3$. However, the detection of star-forming clumps and the definition of clumpy SFGs also vary in both studies and the $f_{\mathrm{clumpy}}$ values are difficult to directly compare with our results.

Our observed clumpy fraction for SFGs at redshifts $z\leq 0.32$ and with $M_\star > 10^9\,M_\odot$ follows the trajectory of the cosmic SFR density and is consistent with the expected decline of the clumpy fractions from values observed by \citet{Guo2015} at redshift $0.5 \leq z \leq 3.0$. This is shown in Figure \ref{fig:hsc_phys_props_galaxy_clumpy_fraction_mid_z} where we also plot a best fit model of the cosmic SFR density from \citet{Madau2014} that is rescaled for a qualitative comparison with the clumpy fractions at different redshifts. However, the observed clumpy fraction does not continue to decline for redshifts $z < 0.15$. Instead, it appears that $f_{\mathrm{clumpy}}$ has reached a stable value of $\sim 23\%$ for the redshift range probed in this study. This result suggests that clumpy star-formation remains a frequently observable morphological feature in nearby galaxies and is more abundant in low-redshift SFGs than previously assumed.

\begin{figure*}
    \centering
    \includegraphics[width=1\textwidth]{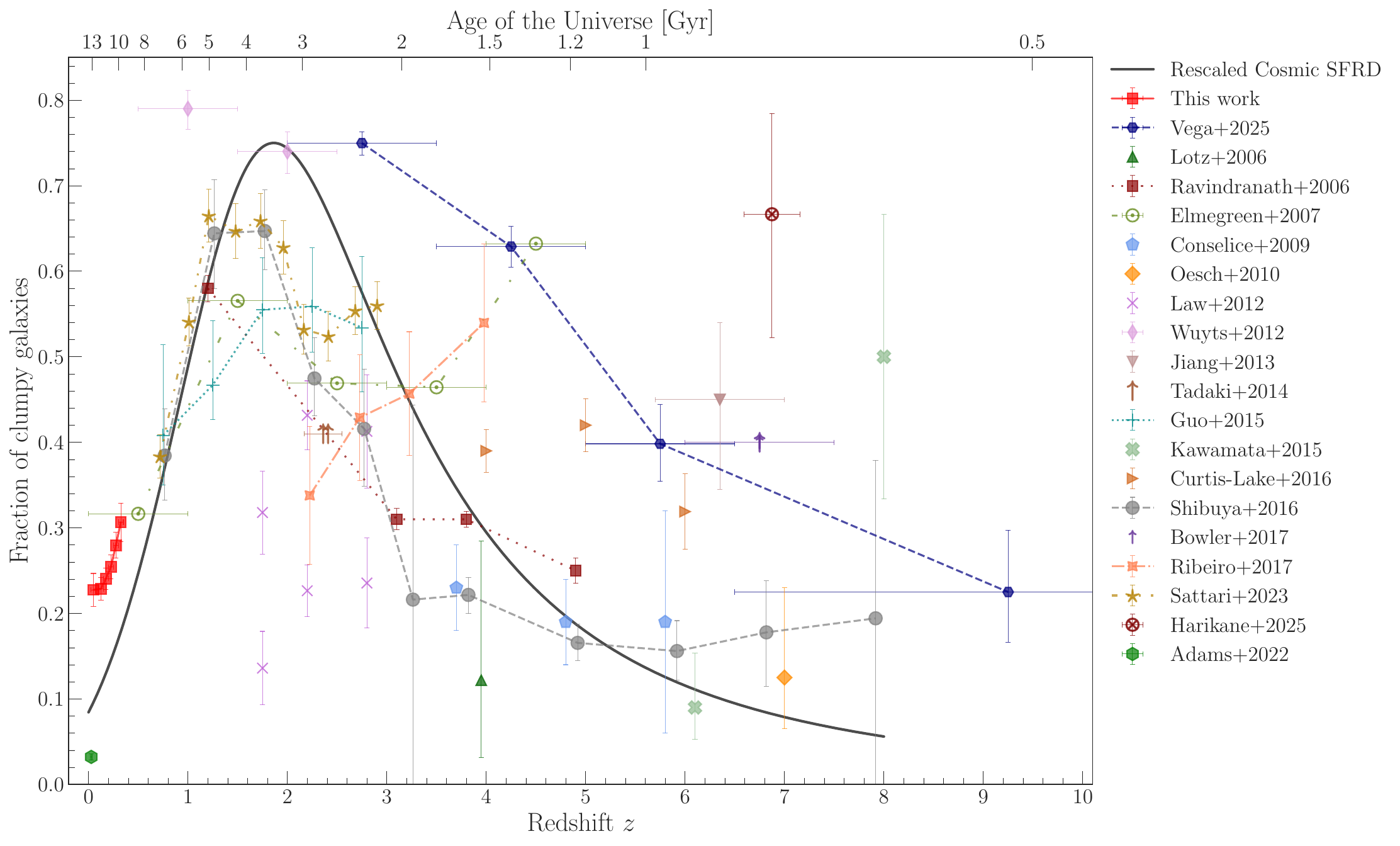}
    \caption[Clumpy fractions of SFGs as a function of redshift.]{Plot of clumpy fractions as a function of redshift published from various authors. This plot has been recreated from \citet[][Fig. 7]{Vega2025}. The clumpy fractions observed for low-redshift galaxies from this study are shown in red. Also shown is the evolution of the cosmic SFR density from \citet{Madau2014} in grey but rescaled to arbitrary units for better comparison. The literature values are from: Vega+2025: \citet{Vega2025}, Lotz+2006: \citet{Lotz2006}, Ravindranath+2006: \citet{Ravindranath2006}, Elmegreen+2007: \citet{Elmegreen2007a}, Conselice+2009: \citet{Conselice2009}, Oesch+2010: \citet{Oesch2010}, Law+2012: \citet{Law2012}, Wuyts+2012: \citet{Wuyts2012}, Jiang+2013: \citet{Jiang2013}, Tadaki+2014: \citet{Tadaki2014}, Guo+2015: \citet{Guo2015}, Kawamata+2015: \citet{Kawamata2015}, Curtis-Lake+2016: \citet{CurtisLake2016}, Shibuya+2016: \citet{Shibuya2016}, Bowler+2017: \citet{Bowler2016}, Ribeiro+2017: \citet{Ribeiro2017}, Sattari+2023: \citet{Sattari2023}, Harikane+2025: \citet{Harikane2025} and Adams+2022: \citet{Adams2022}.}
    \label{fig:hsc_phys_props_galaxy_clumpy_fraction_high_z}
\end{figure*}

Historically, the fraction of clumpy galaxies was mainly determined using studies of high- and very high-redshift galaxies ($1 \lesssim z \lesssim 8$). Similar studies focusing on the nearby Universe at redshifts $z \lesssim  0.5$ are less common. As a qualitative comparison, we added our measurements of the clumpy fractions in the low-redshift regime to the clumpy fractions observed for the high-redshift regimes and specifically over the epoch of peak cosmic star-formation in Figure \ref{fig:hsc_phys_props_galaxy_clumpy_fraction_high_z}. Our findings are consistent with what would be expected for the fraction of clumpy SFGs at low redshifts and align well with the assumed trend of the cosmic SFR density. 

\subsection{Dependence on host galaxy properties of the clumpy fraction}\label{sec:clumpy_fraction_host}
\begin{figure}
    \centering
    \subfloat[\centering Clumpy fraction per redshift bin. \label{fig:hsc_phys_props_galaxy_clumpy_fraction_mass_a}]{{\includegraphics[width=0.7\columnwidth]{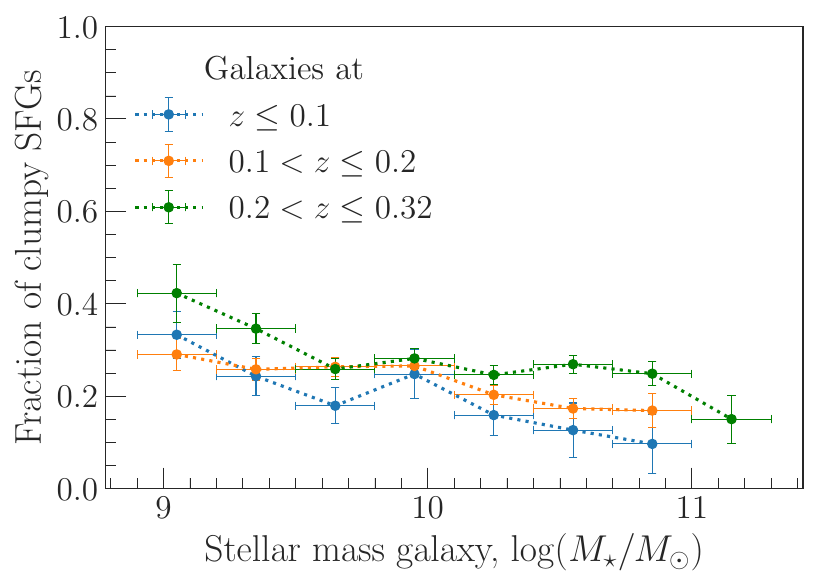} }}
    \\
    \subfloat[\centering Clumpy fraction (all galaxies, $z \leq 0.32)$. \label{fig:hsc_phys_props_galaxy_clumpy_fraction_mass_b}]{{\includegraphics[width=0.7\columnwidth]{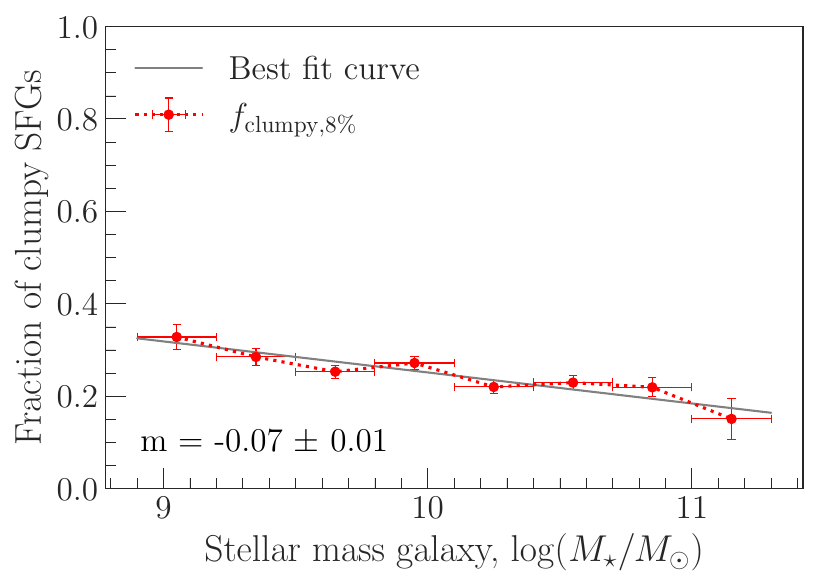} }}
    \caption[Incompleteness-corrected clumpy fraction as a function of the galaxy's stellar mass.]{Incompleteness-corrected clumpy fraction as a function of the galaxy's stellar mass. The plots show the clumpy fractions for clumps defined by a clump-galaxy flux ratio $F_{u,\mathrm{cl}}/F_{u,\mathrm{gal}} \geq 0.08$ for three different redshift bins (a) and overall (b). A linear model is fitted to the values of all galaxies and $m= \dd \log(M_\star)/\dd (f_{\mathrm{clumpy,8\%}})$ is indicated in panel (b). Only galaxies from the mass-complete sample are shown. Error bars show the standard error for the clumpy fractions and the bin range for the galaxy stellar mass.}
    \label{fig:hsc_phys_props_galaxy_clumpy_fraction_mass}
\end{figure}

The individual clumpy fractions for each of the three analysed mass bins differ but roughly follow a similar trend for all mass bins, indicating a dependence of $f_{\mathrm{clumpy}}$ on the stellar mass of the host galaxy. This is presented in Figure \ref{fig:hsc_phys_props_galaxy_clumpy_fraction_mass}, where we plot the clumpy fraction as a function of stellar mass for different redshift bins (Fig. \ref{fig:hsc_phys_props_galaxy_clumpy_fraction_mass_a}) and combined (Fig. \ref{fig:hsc_phys_props_galaxy_clumpy_fraction_mass_b}). No noticeable dependence on redshift is evident from Figure \ref{fig:hsc_phys_props_galaxy_clumpy_fraction_mass_a}. Instead, the clumpy fraction declines with increasing stellar mass of the host galaxy for all galaxies in the mass-complete sample that have redshift $z \leq 0.32$ (Fig. \ref{fig:hsc_phys_props_galaxy_clumpy_fraction_mass_b}). This is also consistent to findings from \citet{Murata2014,Guo2015,Sattari2023}, although the clumpy fractions of the low-redshift galaxies analysed here are lower and the decline appears to be less steep than what has been observed from galaxies at higher redshifts. For example, \citet{Guo2015} observed a decline of $f_{\mathrm{clumpy}}$ from $\sim$60\% to $\sim$30\% for galaxies with stellar masses of $\log(M_\star/M_\odot) = 9.0$ to $11.4$ at redshift $0.5 < z < 3.0$, whereas we observe a decline from $\sim$35\% to $\sim$20\% over the same range of galaxy stellar masses. The continuous decline of $f_{\mathrm{clumpy}}$ with increasing stellar mass of the host galaxy suggests that the disks of more massive galaxies are less turbulent, which means that clumps are less likely to form by VDI. This could be due to the stabilising effect of more massive dark matter haloes in low-redshift SFGs \citep[][for example]{Jog2014}. 

\begin{figure}
    \centering
    \subfloat[\centering Clumpy fraction per redshift bin. \label{fig:hsc_phys_props_galaxy_clumpy_fraction_ssfr_a}]{{\includegraphics[width=0.7\columnwidth]{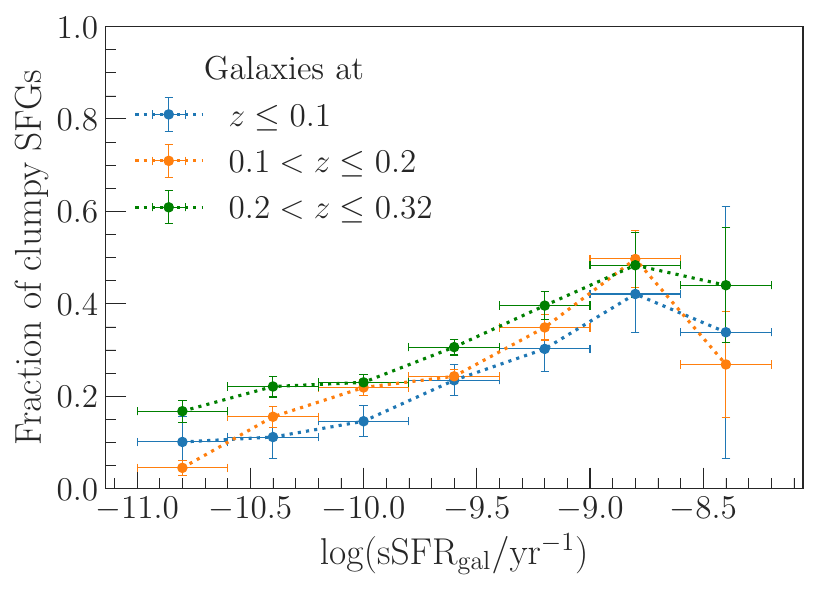} }}
    \\
    \subfloat[\centering Clumpy fraction (all galaxies, $z \leq 0.32)$. \label{fig:hsc_phys_props_galaxy_clumpy_fraction_ssfr_b}]{{\includegraphics[width=0.7\columnwidth]{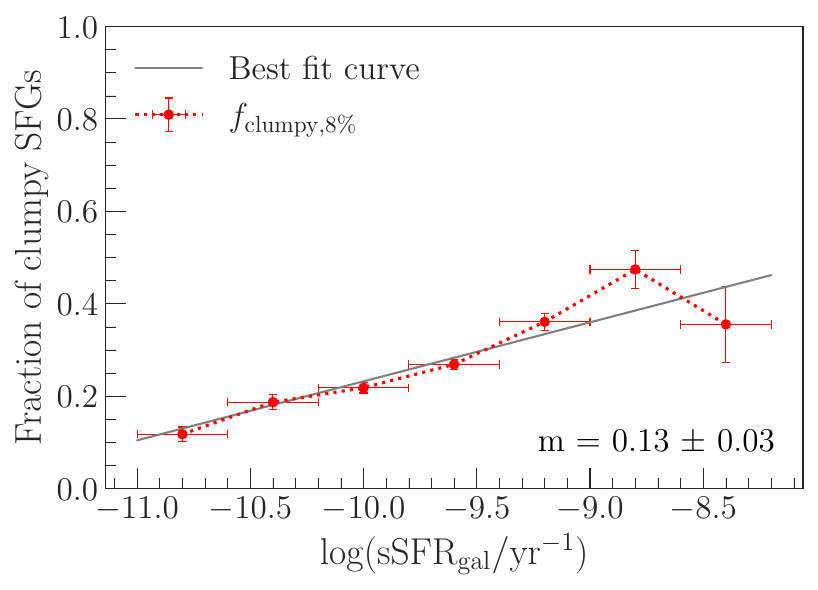} }}
    \caption[Incompleteness-corrected clumpy fraction as a function of the galaxy's sSFR.]{Incompleteness-corrected clumpy fraction as a function of the galaxy's sSFR. The plots show the clumpy fractions for clumps defined by a clump-galaxy flux ratio $F_{u,\mathrm{cl}}/F_{u,\mathrm{gal}} \geq 0.08$ for three different redshift bins (a) and overall (b). A linear model is fitted to the values of all galaxies and $m= \dd (\log \mathrm{sSFR}_\mathrm{gal})/\dd (f_{\mathrm{clumpy,8\%}})$ is indicated in the panel (b). Only galaxies from the mass-complete sample are shown. Error bars show the standard error for the clumpy fractions and the bin range for the galaxy sSFR.}
    \label{fig:hsc_phys_props_galaxy_clumpy_fraction_ssfr}
\end{figure}

An observable increase in $f_{\mathrm{clumpy}}$ with increasing sSFR of the host galaxy is expected as clumps are regions of intense star-formation. This increase is evident in Figure \ref{fig:hsc_phys_props_galaxy_clumpy_fraction_ssfr} where we show the clumpy fractions for the mass-complete sample as a function of $\log(\mathrm{sSFR}_\mathrm{gal} / \mathrm{yr}^{-1})$ for different redshift bins (Fig. \ref{fig:hsc_phys_props_galaxy_clumpy_fraction_ssfr_a}) and for all galaxies combined (Fig. \ref{fig:hsc_phys_props_galaxy_clumpy_fraction_ssfr_b}). A similar relation between the sSFR and the clumpy fraction of SFGs was also observed by \citet{Murata2014,Shibuya2016}. The increase of $f_{\mathrm{clumpy}}$ with increasing sSFR is independent of redshift (Fig. \ref{fig:hsc_phys_props_galaxy_clumpy_fraction_ssfr_a}).

\begin{figure}
    \centering
    \subfloat[\centering Clumpy fractions per sSFR and stellar mass bin for $F_{u,\mathrm{cl}}/F_{u,\mathrm{gal}} \geq 0.08$. \label{fig:hsc_phys_props_galaxy_clumpy_fraction_ssfr_mass_a}]{{\includegraphics[width=1\columnwidth]{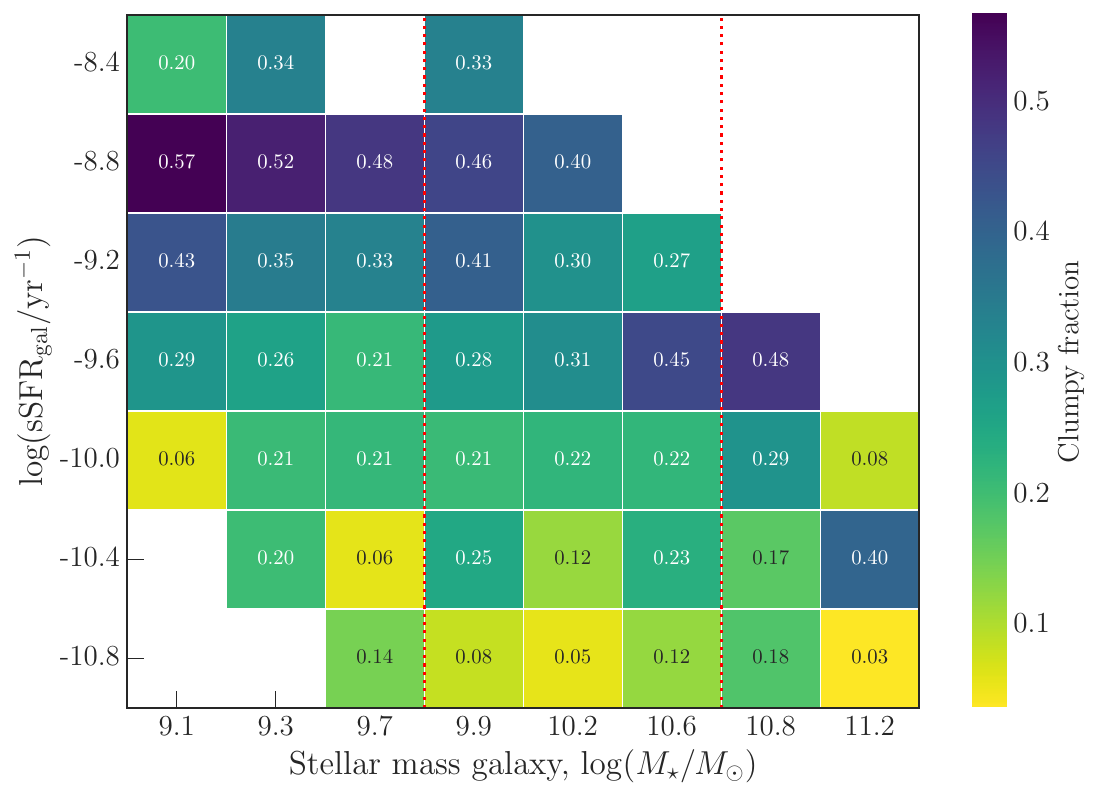} }}
    \\
    \subfloat[\centering Count of SFGs per sSFR and stellar mass bin. \label{fig:hsc_phys_props_galaxy_clumpy_fraction_ssfr_mass_b}]{{\includegraphics[width=1\columnwidth]{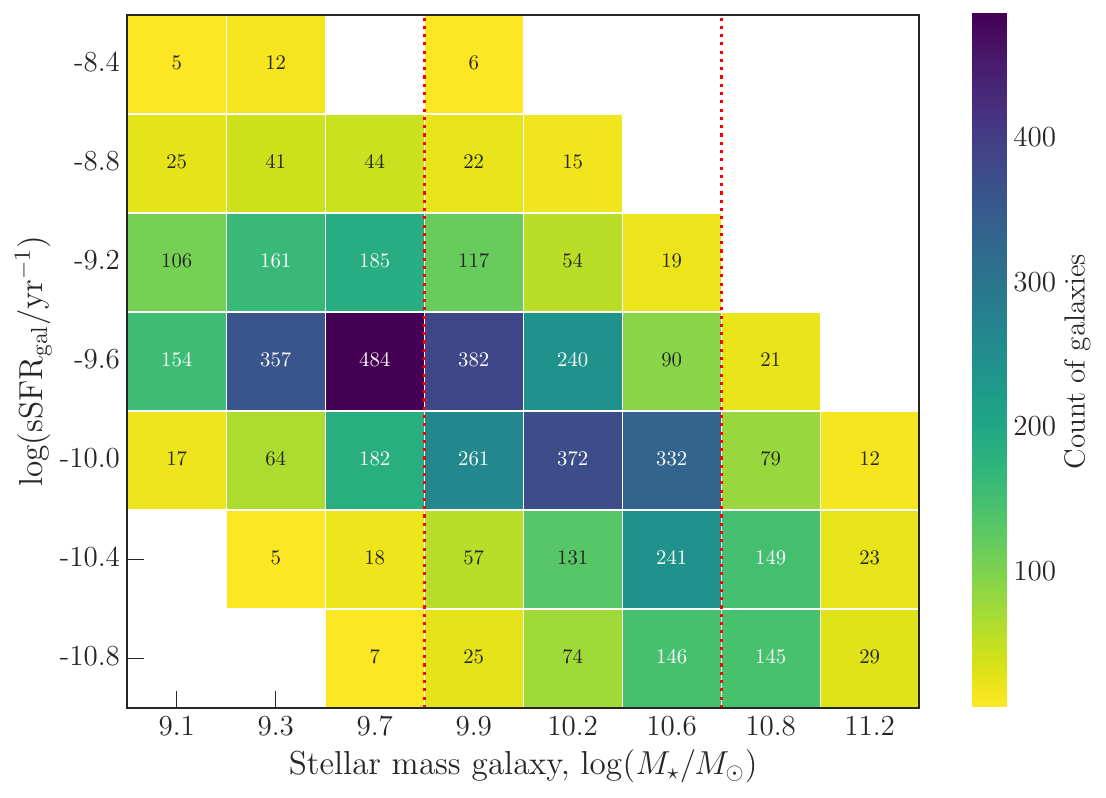} }}
    \caption[Incompleteness-corrected clumpy fraction and count of SFGs per sSFR and stellar mass bin of the host galaxies.]{Incompleteness-corrected clumpy fraction for $F_{u,\mathrm{cl}}/F_{u,\mathrm{gal}} \geq 0.08$ and count of SFGs per sSFR and stellar mass bin of the host galaxies. Plot (a) shows the clumpy fractions for clumps defined by a clump-galaxy flux ratio $F_{u,\mathrm{cl}}/F_{u,\mathrm{gal}} \geq 0.08$ and plot (b) the corresponding count of SFGs in each bin. Only the mass-complete sample is shown and bins with less than five SFGs have been omitted for clarity. The dotted red vertical lines show the stellar mass limits of the low-mass galaxy bin ($9.0 \leq\log(M_\star/M_\odot) < 9.8$), the medium-mass galaxy bin ($9.8 \leq\log(M_\star/M_\odot) < 10.6$) and the high-mass galaxy bin ($10.6 \leq\log(M_\star/M_\odot) < 11.4$).}
    \label{fig:hsc_phys_props_galaxy_clumpy_fraction_ssfr_mass}
\end{figure}

Figure \ref{fig:hsc_phys_props_galaxy_clumpy_fraction_ssfr_mass} shows $f_{\mathrm{clumpy}}$ and the corresponding number of SFGs in a sSFR/$M_\star$ diagram for stellar mass and sSFR bins with the same size as in Figure \ref{fig:hsc_phys_props_galaxy_clumpy_fraction_mass} and \ref{fig:hsc_phys_props_galaxy_clumpy_fraction_ssfr}. The clumpy fractions tend to increase from $\sim$15\% for galaxies at the high-mass and low-sSFR end to $\sim$50\% for galaxies at the low-mass and high-sSFR end of the mass-complete sample (from bottom right to top left in Fig. \ref{fig:hsc_phys_props_galaxy_clumpy_fraction_ssfr_mass_a}). 

\subsection{Comparison of different measurements of the clumpy fraction}\label{sec:clumpy_fraction_diff_defs}
\begin{figure}
    \centering
    \includegraphics[width=0.7\columnwidth]{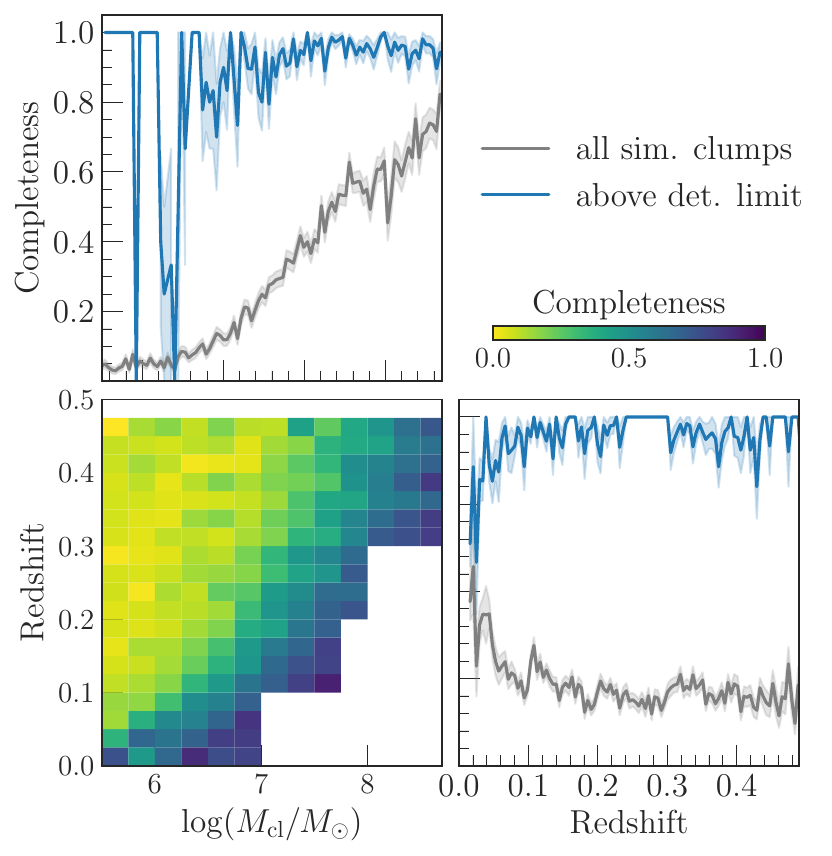}
    \caption[Detection completeness of the simulated clumps as a function of redshift and clump stellar mass.]{Detection completeness measured on the full sample of simulated clumps as a function of redshift and clump stellar mass. The completeness over the range of the single variables are shown by the line plots in blue for detected simulated clumps with measured fluxes above the average $5\sigma$ point-source depth limits of the filter bands and in grey without any flux cuts. The $1\sigma$ errors are shown as shaded areas. The 2-dimensional histogram shows the completeness distribution for all detected simulated clumps without any flux cuts. The stellar masses of the simulated clumps were cut at an upper limit depending on the redshift of the host galaxy to specifically validate completeness limits extending into low stellar mass ranges.}
    \label{fig:sim_clumps_det_completeness_mass_z}
\end{figure}

So far, our measurements of the clumpy fraction and their dependence on the host galaxy properties are based on clumps that are defined by a clump-galaxy flux ratio $F_{u,\mathrm{cl}}/F_{u,\mathrm{gal}} \geq 0.08$. However, \citet{HuertasCompany2020} find that the clumpy fraction increases with stellar mass and that there is no significant dependence on the sSFR of the host galaxy. The authors argue that these contradicting relations are likely due to a different clump definition based on a clump stellar mass threshold of $M_{\mathrm{cl}} \geq 10^7\,M_\odot$ instead of the clump-galaxy flux ratio used by other authors. In order to test how our observation would compare to the findings from \citet{HuertasCompany2020}, we used the estimate of the stellar mass of our clump sample that we obtained through SED fitting (Popp et al., 2026c, submitted) to apply the same mass threshold, i.e. $M_{\mathrm{cl}} \geq 10^7\,M_\odot$. Again, only clumps with stellar masses of $M_{\mathrm{cl}} \geq 10^7\,M_\odot$ that have measured fluxes above the average $5\sigma$ point-source depth limits of the filter bands were used, for which the detection completeness of our clump detector is $>0.9$ (Fig. \ref{fig:sim_clumps_det_completeness_mass_z}). We also corrected the observed clumpy fractions using a similar approach to the one we applied to $f_{\mathrm{clumpy}}$ based on the clump-galaxy \textit{u}-band flux ratios (Appendix \ref{sec:clumpy_fraction_incomplete_calc}). We note, however, that some clumps with $M_{\mathrm{cl}} \geq 10^7\,M_\odot$ and fluxes below the filter band-specific detection limits are likely to be missed by our clump detector (see Fig. \ref{fig:sim_clumps_det_completeness_mass_z}). The completeness of our detections also decreases with increasing redshift of the host galaxies for those clumps. Therefore, the clumpy fractions, which we measured from the magnitude-limited clump sample and that are based on a stellar mass threshold of $M_{\mathrm{cl}} \geq 10^7\,M_\odot$, represent only a lower limit of the true clumpy fractions.

\begin{figure}
    \centering
    \includegraphics[width=0.7\columnwidth]{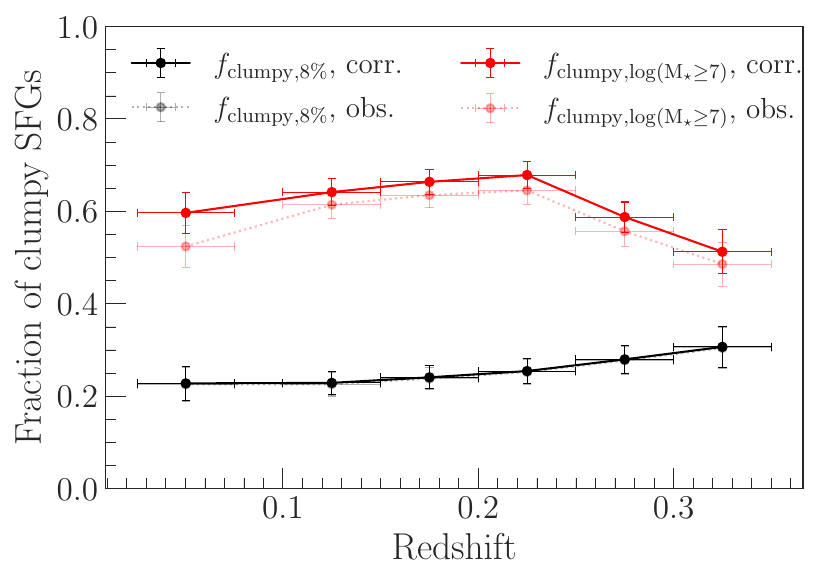}
    \caption[Clumpy fraction of SFGs as a function of redshift and for different clump definitions.]{Clumpy fraction of SFGs as a function of redshift and for different clump definitions. The clumpy fractions defined by a clump-galaxy flux ratio $F_{u,\mathrm{cl}}/F_{u,\mathrm{gal}} \geq 0.08$ ($f_{\mathrm{clumpy,8\%}}$) are plotted in black and the those defined by a clump stellar mass threshold of $M_{\mathrm{cl}}\geq 10^7 M_\odot$ ($f_{\mathrm{clumpy,\log(M_\star\geq7)}}$) in red. Incompleteness-corrected clumpy fractions are shown with opaque colours and observed clumpy fractions with half-transparent colours. Error bars show the standard error for the clumpy fractions and the bin range for redshift.}
    \label{fig:hsc_phys_props_galaxy_clumpy_fraction_low_z_all}
\end{figure}

The clumpy fractions $f_{\mathrm{clumpy,\log(M_\star\geq7)}}$ we observe using the mass-based definition are $\sim$60\% (Fig. \ref{fig:hsc_phys_props_galaxy_clumpy_fraction_low_z_all}), which is $\sim$2 to 3 times higher than $f_{\mathrm{clumpy,8\%}}$ that is measured for clumps defined by a relative flux criterion. The much higher clumpy fractions of the SFGs that are observed from the mass-based clump definition and compared to those that are based on the clump-galaxy flux ratio, also suggest that many massive clumps exist without being very luminous in the UV- or NUV/\textit{u}-band. Star-forming clumps appear to be a common morphological feature in SFGs at $z\lesssim 0.3$ and clumpy star-formation is not only dominant in high-redshift galaxies ($1 < z < 3$) but may also play a significant role in low-redshift galaxies.

\begin{figure}
    \centering
    \subfloat[\centering Clumpy fractions as a function of $M_\star$. \label{fig:hsc_phys_props_galaxy_clumpy_fraction_mass_huertas_a}]{{\includegraphics[width=0.7\columnwidth]{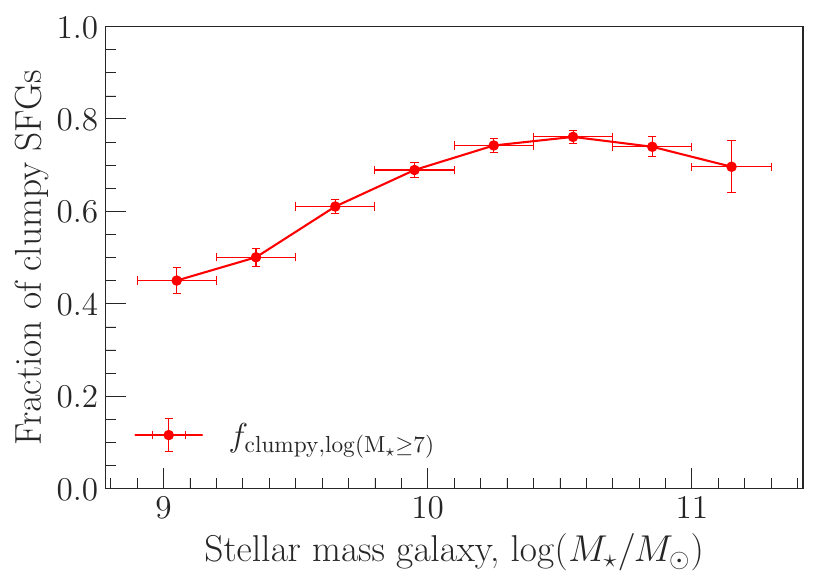} }}
    \\
    \subfloat[\centering Clumpy fractions as a function of sSFR. \label{fig:hsc_phys_props_galaxy_clumpy_fraction_mass_huertas_b}]{{\includegraphics[width=0.7\columnwidth]{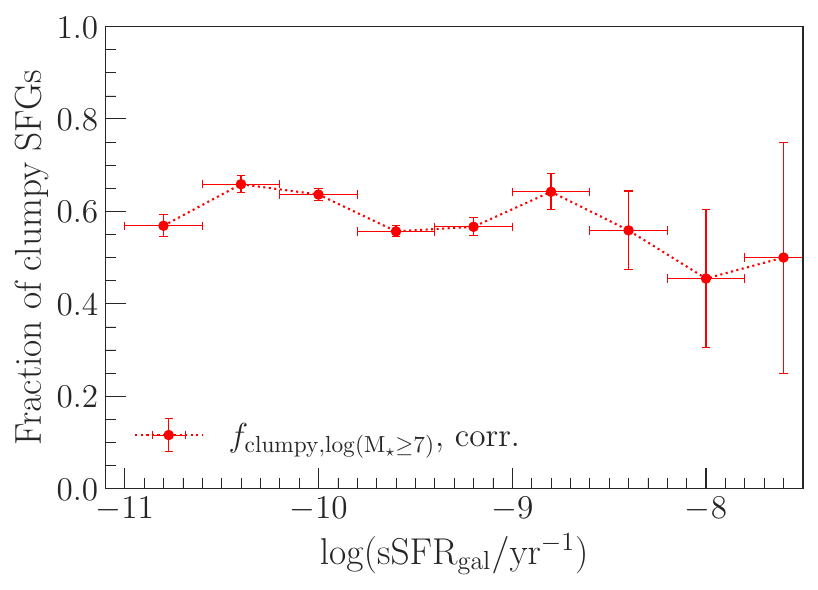} }}
    \caption[Incompleteness-corrected clumpy fraction as a function of the galaxy's stellar mass and sSFR for clumpy SFGs with at least one off-centre clump $M_{\mathrm{cl}} \geq 10^7 M_\odot$.]{Incompleteness-corrected clumpy fraction based on the definition by \citet{HuertasCompany2020} as a function of the galaxy's stellar mass and sSFR. Clumpy galaxies have at least one off-centre clump with $M_{\mathrm{cl}} \geq 10^7 M_\odot$. The clumpy fraction is shown as a function of the galaxy's stellar mass in (a) and the galaxy's sSFR in (b). Error bars show the standard error for the clumpy fractions and the bin range for the galaxy stellar mass and sSFR.}
    \label{fig:hsc_phys_props_galaxy_clumpy_fraction_mass_huertas}
\end{figure}

Compared to the observations from \citet{HuertasCompany2020}, $f_{\mathrm{clumpy},\log(M_\star)\geq7}$ is also higher by a factor of $\sim$2 to 3, but shows a similar trend in that the clumpy fraction increases until $M_{\mathrm{gal}} \lesssim 10^{10.5}\,M_\odot$ and then declines again for more massive galaxies (Fig. \ref{fig:hsc_phys_props_galaxy_clumpy_fraction_mass_huertas_a}). The increase of the clumpy fraction appears to be linked to the mass-based definition of a clump. Less massive galaxies are less likely to host very massive clumps that would contain a considerable fraction of the total galaxy stellar mass.

In contrast with the continuous decline of the clumpy fraction with increasing mass as seen from the flux ratio-based clump definition (Fig. \ref{fig:hsc_phys_props_galaxy_clumpy_fraction_mass_b}), the decline of the clumpy fraction for only high-mass galaxies ($M_{\mathrm{gal}} \gtrsim 10^{10.5}\,M_\odot$) using the mass-based clump definition might indicate a change towards a less turbulent galactic disk environment. A more stable environment might result in either the formation of fewer very massive clumps and/or star-formation that is less concentrated in clump regions with $M_{\mathrm{cl}} \geq 10^7 M_\odot$ and more distributed over wider regions of the galaxy disk. Also, the increase of $f_{\mathrm{clumpy,8\%}}$ with increasing sSFR seen in Figure \ref{fig:hsc_phys_props_galaxy_clumpy_fraction_ssfr} is less obvious if the clump definition is changed to $f_{\mathrm{clumpy},\log(M_\star)\geq7}$ (Fig. \ref{fig:hsc_phys_props_galaxy_clumpy_fraction_mass_huertas_b}). The clumpy fraction fluctuates between $\sim 50$-70\% and shows no significant dependence on the sSFR of the host galaxy, which is in better agreement with the findings from \citet{HuertasCompany2020}.

\begin{figure}
    \centering
    \includegraphics[width=1\columnwidth]{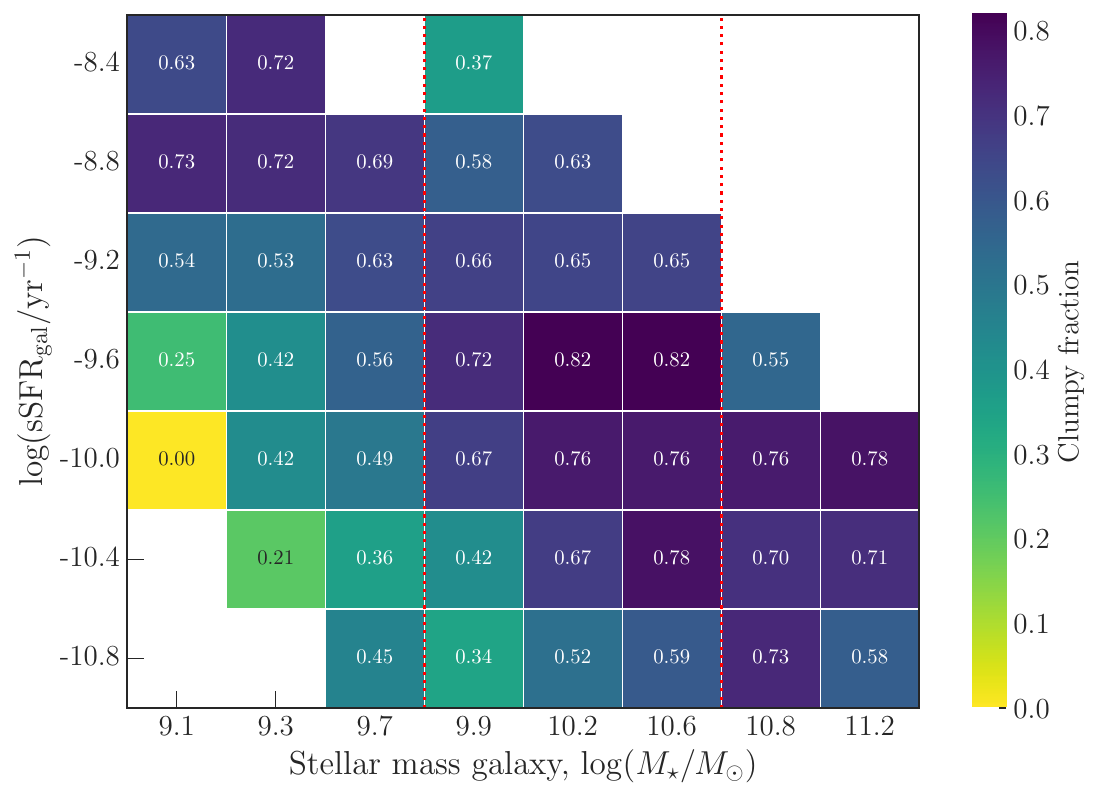}
    \caption[Incompleteness-corrected clumpy fraction for clumps with $M_{\mathrm{cl}} \geq 10^7 M_\odot$ per sSFR and stellar mass bin of the host galaxies.]{Similar to Figure \ref{fig:hsc_phys_props_galaxy_clumpy_fraction_ssfr_mass_a} but showing the incompleteness-corrected clumpy fraction for clumps with $M_{\mathrm{cl}} \geq 10^7 M_\odot$ per sSFR and stellar mass bin of the host galaxies.}
    \label{fig:hsc_phys_props_galaxy_clumpy_fraction_ssfr_mass_huertas}
\end{figure}

Figure \ref{fig:hsc_phys_props_galaxy_clumpy_fraction_ssfr_mass_huertas} shows the incompleteness-corrected $f_{\mathrm{clumpy},\log(M_\star)\geq7}$ of the galaxies from the HSC-SSP sample in a sSFR/$M_\star$ diagram for SFGs in the same stellar mass and sSFR bins that were used in Figure \ref{fig:hsc_phys_props_galaxy_clumpy_fraction_mass_huertas}. Here, the clumpy fractions appear to first increase with increasing stellar mass and sSFR of the host galaxy until they remain at a level of $\sim$50-70\%. This could possibly be due to a change of the turbulent disk environment that is not only determined by the stellar mass but also by the sSFR of the host galaxies. Initially, the formation of large clumps with $M_{\mathrm{cl}} \geq 10^7\,M_\odot$ is likely limited by the stellar mass of the galaxy such that only galaxies with more intense star-formation are capable of forming massive clumps, which contain a considerable fraction of the total galaxy stellar mass. With increasing galaxy mass, the disk environment may become more stable with a less turbulent environment leading to the formation of fewer massive clumps by VDI in high-mass galaxies ($M_{\mathrm{gal}} \gtrsim 10^{10.5}\,M_\odot$). This suppression of clump formation may not be complete and clumps with $M_{\mathrm{cl}} \geq 10^7 M_\odot$ can still form in galaxies with high sSFRs even if those galaxies are very massive. The most massive galaxies might also already be in a transitional phase to becoming quiescent after most of the gas that fuels star-formation has been either consumed or driven out of the galaxy.

The differences in the relations between the clumpy fraction and the physical properties of the host galaxies that were observed in this study, and the relations published by various other authors, are mainly due to incompatible definitions of a star-forming clump. Whereas a definition based on the UV/NUV/\textit{u}-band clump-galaxy flux ratio threshold is focused on young star-forming regions with intensive star-forming activity, a definition based on a clump stellar mass threshold also takes into account older and less UV/\textit{u}-bright clumps. This is particularly true if the detection method uses imaging data that also covers the optical and NIR wavelength ranges. For example, the clump definition from \citet{HuertasCompany2020} results in much higher clumpy fractions than the clump definition from \citet{Guo2015}, if applied to our observations (Fig. \ref{fig:hsc_phys_props_galaxy_clumpy_fraction_low_z_all}). This complicates the direct comparison of the derived clumpy fractions for SFGs from different studies, particularly due to the fact that the studies have probed different redshift ranges.

A more precise analysis of the observed relations of the clumpy fraction with the host galaxy properties requires more detailed data on the clumps and the host galaxies. Many other parameters and processes are not accounted for, e.g. AGN feedback, possible mergers, the intergalactic environment of the host galaxy and the stability criteria for gravitationally bound or unbound clumps in gas-rich turbulent disks. Upcoming research projects will make use of imaging data from \textit{Euclid} with higher spatial resolution and sensitivity that will cover a much larger sample of galaxies at greater depths. Analysing the clumpy fractions for galaxies that are mass-complete to $\ll 10^9M_\odot$ over a redshift range of $0 < z < 0.5$ will help to describe the relations of $f_{\mathrm{clumpy}}$ to the host galaxy properties in better detail. \textit{Euclid} will also acquire spectroscopic data for a large set of observed galaxies, which will provide more detailed information about the star-forming processes operating in star-forming clumps.

\section{Summary and conclusions}\label{sec:conclusion}
In this work, we identified clumps in a mass-complete sample of 5,395 low-redshift SFGs ($0.0005 \leq z \leq 0.32$) observed by CLAUDS and HSP-SSP in the XMM-LSS, E-COSMOS and DEEP2-3 fields. Our sample of 12,790 clumps were identified using a DL-based object detection model that was trained to detect objects which are similar to those identified visually by human beings. We carefully validated our detections and aperture photometry measurements of the clumps using a large sample of simulated clumps that were injected into the same galaxy images and showed that detection completeness is high for bright clumps using a clump-galaxy flux ratio thresholds of $F_{u,\mathrm{cl}}/F_{u,\mathrm{gal}} \geq 0.08$. We measured the clumpy fraction $f_{\mathrm{clumpy}}$ based on a clump definition that requires a clump-galaxy flux ratio in the CLAUDS \textit{u}-band of $\geq8\%$ and also compare our results to recent observations of clumpy galaxies made by other studies.

The key results are summarised in the following points:
\begin{enumerate}
  \item The estimated and incompleteness-corrected fraction of SFGs hosting at least one off-centre clump increases from $\sim 23\%$ to $\sim 31\%$ over the redshift range of $0.005 \leq z \leq 0.32$. These clumpy fractions are significantly increased with respect to previous observations of comparable low-redshift SFGs.
  \item The clumpy fraction tends to decrease with increasing stellar mass of the host galaxy while increasing with increasing sSFR of the host galaxy at fixed redshift.
  \item The redshift evolution of $f_{\mathrm{clumpy}}$ is consistent with a low-redshift extrapolation of the clumpy fraction that is measured using high-redshift observations and is expected to decline from $\sim$60\% at $z\sim 1$-2.
  \item We observe a significantly increased clumpy fraction of $f_{\mathrm{clumpy}} \sim 60\%$ for a changed clump definition that is based on a clump stellar mass threshold of $M_{\mathrm{cl}} \geq 10^7 M_\odot$. In contrast to a clumpy fraction based on a relative flux criterion, the mass-based clumpy fraction tends to increase with stellar mass of the host galaxy and does not show an observable dependence on the sSFR of the host galaxy.
\end{enumerate}

\section*{Acknowledgements}
We would like to thank the anonymous referee for their valuable comments and insight, which improved the quality of this paper.

JJP acknowledges funding from the Science and Technology Facilities Council (STFC) Grant Code ST/X508640/1. HD and SS acknowledge funding via the ELSA project. ``ELSA: Euclid Legacy Science Advanced analysis tools'' (Grant Agreement no. 101135203) is funded by the European Union. Views and opinions expressed are however those of the author(s) only and do not necessarily reflect those of the European Union or Innovate UK. Neither the European Union nor the granting authority can be held responsible for them. UK participation is funded through the UK Horizon guarantee scheme under Innovate UK grant 10093177. LFF acknowledges partial support from NASA awards 80NSSC24K1277 and 80NSSC20M0057.

This research made use of the open-source Python scientific computing ecosystem, including \textsc{NumPy} \citep{Harris2020}, \textsc{Matplotlib} \citep{Hunter2007}, \textsc{seaborn} \citep{Waskom2021} and \textsc{Pandas} \citep{McKinney2010}. This research made use of \textsc{Astropy}, a community-developed core Python package for Astronomy \citep{astropy2022} and the \textsc{Photutils} Python package \citep{LarryBradley2025}.

\section*{Data Availability}
The training data and Python code examples for the FRCNN models and the adjusted feature extraction backbone are available from \citet{Popp2025} and a public Github repository\footnote{\url{https://github.com/ou-astrophysics/Zoobot-for-image-segmentation-and-object-detection}}. The catalogue of star-forming clumps, including the measured photometry and estimated physical properties, is available from \citet{Popp2026}.

\section*{Conflicts of Interest}
The authors declare no conflict of interest.



\bibliographystyle{mnras}
\bibliography{LIB_Clumpy_Galaxies} 




\appendix
\section{Clump detection}\label{sec:frcnn_development}
Object detection is one of many technologies used in computer vision and its main application is in detecting and recognising instances of \textit{semantic} objects, i.e. objects of meaningful physical origin, in images or videos \citep[e.g.][]{Dasiopoulou2005}. Object detection algorithms use Machine Learning (ML) or Deep Learning (DL) to produce automatic detections of all instances of multiple objects in an image but also assigns a label to each instance found. Each detected instance of a specific object is marked with a tightly cropped bounding box centred on the instance.

In this paper we train a DL-based object detection model, specifically a version of the Faster R-CNN (FRCNN) architecture proposed by \citet{Ren2015}. FRCNN models comprise three components: (1) a CNN that is used as a `backbone' to extract spatial hierarchies of patterns or features from an input image and that are then used as input to two separate sub-networks, (2) a Region Proposal Network (RPN) and (3) a detector network. The RPN identifies regions in the image that are likely to contain objects. It sets anchor points at every pixel location of the output feature map of the feature extracting backbone and places at each anchor point position a set of $k$ anchor boxes with default sizes and aspect ratios. It then optimises the size of the initial anchor boxes depending on the overlap with the ground-truth object boxes from the training set and generates a prediction score (`objectness') for the two generic classes, `object' and `background'. The second sub-network, the detector network, is then used to classify the contents of the proposed regions of class `object' into one of the $n$ final object categories using the corresponding features for those parts of the image that were extracted by the backbone CNN. It also further refines the predicted bounding boxes. The final output of the FRCNN model is a collection of rectangular bounding boxes identifying groups of pixels in the image that contain objects and a classification identifying the type of object that each box contains.

To train our models, we manually labelled and marked clumps and potential contaminants in a subset of our galaxy images. Each galaxy was presented with four different images to help with the visual identification of potential clumps: (1) a RGB-composite image using HSC-colours, (2) a RGB-composite image with a different scaling of the \textit{g}-band to emphasise star-forming regions, (3) an \textit{u}-band image with simple asinh-stretch and (3) the same \textit{u}-band image with asinh-stretch but where the maximum level of the pixel values is cut at the 99th percentile to emphasise low-surface-brightness features of the target galaxy. We marked the location of clumps with small boxes in the images and also marked other contaminating features visible in or around the target galaxy. These contaminants include:
\begin{enumerate}
    \item \textbf{`odd' clumps} - objects that appear like clumps but are likely image artifacts or other anomalies,
    \item \textbf{foreground stars} - stars in the Milky Way that blend into the target galaxy,
    \item \textbf{fore-/background galaxies} - smaller galaxies in front of the target galaxy, distant galaxies behind the target galaxies or satellite galaxies of the target galaxy,
    \item \textbf{(secondary) bulges} - bright bulges of close-by galaxies or merging galaxies (also used to correct the initial clump predictions if the central bulge of the target galaxy is marked as a clump).
\end{enumerate}

Our training data consists of 16,165 annotations in 3,198 galaxies that were identified from \textit{u}-band and RGB-composite images. However, we expected that additional information from which our model could `learn' to identify clumps are also provided by the \textit{grizy}-filter band data separately. Here, `learning' translates to adjusting the weights of our backbone CNN (and also the connected RPN and detector network) while training simultaneously on the multi-band data from CLAUDS and HSC-SSP. Therefore, we projected our annotations onto each greyscale image of the six \textit{ugrizy}-filter band science images of our training galaxies to create the six channel input required for our object detection model \citep[see][for details]{Popp2026b}. Subsequent tests comparing models that use less than the six imaging channels as input with models that use all six channels showed that purity and completeness of the model detections are increased by $\sim15$ to 25\% if additional information is made available to the detection model through the additional filter bands. Especially purity is increased as many contaminating objects are correctly identified due to their distinctive signal in the redder filter bands. Also, the added \textit{u}-band data from the CLAUDS survey, which is more sensitive to recent star-formation, increases the number of correct clump detections compared to a model without \textit{u}-band data. We note, however, that a principle bias in our training data cannot be excluded. The FRCNN model is trained to detect objects that are similar to those identified visually by human beings. This bears the risk that the training data might not be complete or contains wrongly labelled objects and any biases would be propagated into the fully trained model.

\begin{figure*}
    \centering
    \includegraphics[width=1.0\textwidth]{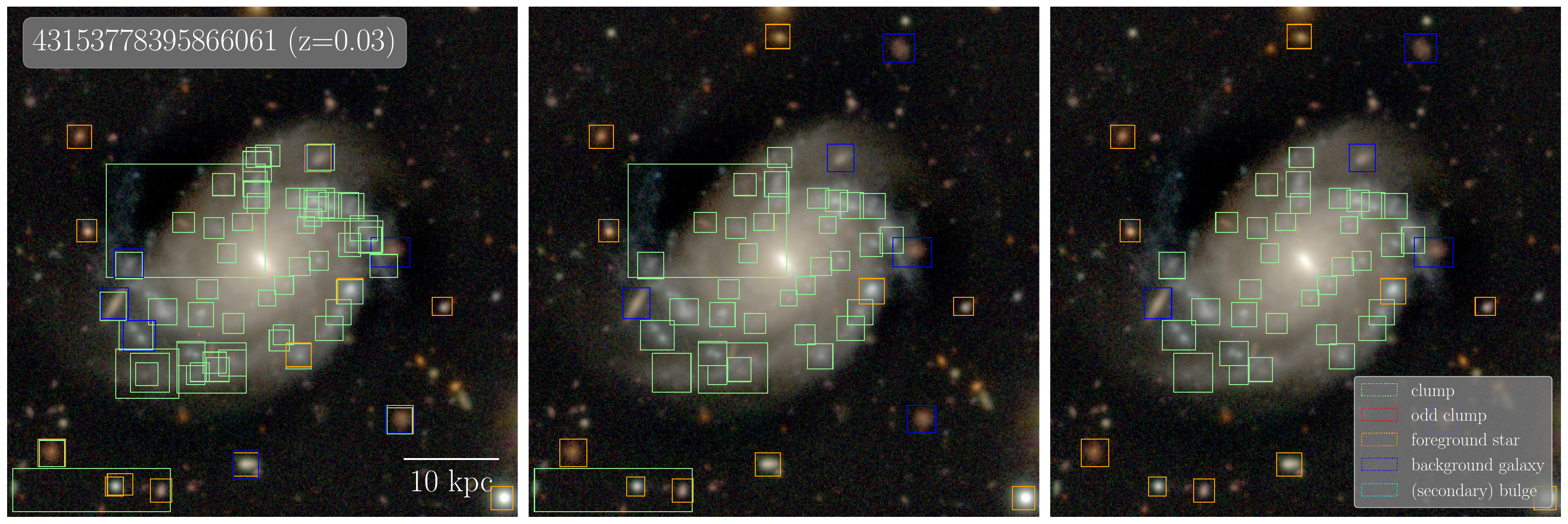}
    \caption[Postprocessing of the FRCNN model detections.]{Example galaxy (object 43153778395866061, z=0.03) showing the different postprocessing steps applied to the FRCNN model detections \citep{Popp2026b}. The left image shows the raw detection results from the model, the central image the detection results after the non-maximum suppression process was applied and the right image after bounding boxes with size $>7.30\,\mathrm{kpc}$ were removed.}
    \label{fig:hsc_det_over_model_dev_postprocess}
\end{figure*}

After our trained FRCNN model was applied to our full set of galaxy images, we applied a process called non-maximum suppression (NMS) which uses the Jaccard distance  $J(A,B)$ \citep{Jaccard1912} to determine the Intersection over Union (IoU) of the areas $A$ and $B$,
\begin{equation}\label{eq:jaccard}
    J(A, B) = \frac{A \cap B}{A \cup B} \,
\end{equation}
We applied the NMS to all galaxy images using a threshold of $\mathrm{IoU} \geq 0.2$ and kept only those detections and the corresponding object class predictions that have the highest objectness from each subset of overlapping bounding boxes. We show an example of the NMS process in the central image of Figure \ref{fig:hsc_det_over_model_dev_postprocess} in comparison to the raw detections that are shown in the left image of Figure \ref{fig:hsc_det_over_model_dev_postprocess}.

Our detections still contain a few bounding boxes that are too large to mark plausible detections of clumps or any of the contaminating features (e.g. centre image of Figure \ref{fig:hsc_det_over_model_dev_postprocess}). These large bounding boxes were not removed by the NMS as the IoU with much smaller bounding boxes can be lower than our applied threshold. We visually inspected the model detections of a few hundred galaxies and set a maximum bounding box size of $\leq 7.30\,\mathrm{kpc}$, which corresponds to the 95th percentile of the size distribution. All bounding boxes that exceed this size threshold were removed. We further discarded all remaining bounding boxes that fully contain other, smaller bounding boxes which were not removed before as the IoU of the smaller bounding box with the larger bounding box is less than $0.2$. An example of a galaxy with the postprocessed model detections at this stage is shown in the right image in Figure \ref{fig:hsc_det_over_model_dev_postprocess}. Finally, we removed all remaining bounding boxes with clump detections that lie outside the target galaxy's segmentation mask and also discarded those bounding boxes that are close to or coincide with the central bulge of a galaxy.

From the resulting set of bounding boxes that were classified as `clump', we extracted the sub-pixel position of every local flux maximum from the pixels within each bounding box \citep[for details, see][]{Popp2026b}. The local flux maxima or flux peaks mark the positions of the final sample of clump candidates. With the given seeing, features with physical sizes $<1$ kpc can be theoretically resolved in only a small fraction of galaxies at $z\lesssim 0.1$. We therefore treat our detected clump candidates as point-like objects and use their sub-pixel position to centre the aperture we used to measure the clump fluxes (Section \ref{sec:clump_phot}). 

We validated the detection performance of our FRCNN model in two ways. First, the detection results were validated during training on a hold-out sample of the training set. However, this first step only compares the detected bounding boxes with the ground-truth bounding boxes from the training data. Therefore, we used simulated clump-like objects that were injected into the real galaxy images to validate purity and completeness of our detection model. The details of generating the set of 32,241 simulated clumps in 13,789 host galaxies are described in \citet{Popp2026b}. Briefly, we simulated spectra from composite stellar populations (CSPs) assuming a delayed exponentially declining star-formation history (SFH), a \citet{Chabrier2003} initial mass function (IMF) and a \citet{Calzetti2000} dust attenuation curve. The physical properties of each simulated clump were sampled from the distributions listed in Table \ref{tab:hsc_det_over_model_dev_eval_sims_sps_params}.

\begin{table}
	\centering
	\caption[Parameters and sampling distribution for each simulated peak spectrum.]{Parameters and sampling distribution for each simulated peak spectrum.}
    \label{tab:hsc_det_over_model_dev_eval_sims_sps_params} 
	\footnotesize
        \begin{tabular}{lll}
		\hline
		Parameter/Unit & Range & Sampling \\ 
		\hline
		$\tau$/Gyr                       & $[0.1, 0.30]$ & Log-uniform \\ 
        $\log(Z_\star/Z_\odot)$          & $[-2.0, 0.19]$ & Uniform \\
        $\log(Z_{\mathrm{gas}}/Z_\odot)$ & same as $Z_\star$ & Uniform \\
        $\mathrm{A}_\mathrm{V}\,/\,m_{\mathrm{AB}}$ & $[0.0, 4.0]$ & Uniform \\
        $M_\star/M_\odot$                & $[10^4, 10^7]$ for $z\leq0.1$ & Log-uniform \\
                                         & $[10^4, 5\times 10^7]$ for $0.1<z\leq0.1$ & \\
                                         & $[10^4, 10^8]$ for $0.2<z\leq0.3$ & \\
                                         & $[10^4, 5\times 10^8]$ for $z>0.3$ & \\
                                         & max. $0.1\, M_{*,\mathrm{galaxy}}$ & \\
        $t_{\mathrm{age}}/\mathrm{Gyr}$  & $[0.005, 1.0]$ & Uniform \\
        $z$                              & Host galaxy redshift & fixed \\
		\hline
	\end{tabular}
\end{table}

The resulting spectra were redshifted to match the redshift of the host galaxy and integrated over the wavelength range of each individual CLAUDS and HSC filter band to determine the luminosity and colour of each simulated clump which we also treated as point-like objects. We placed up to 30 simulated clumps randomly within the host galaxy while avoiding any overlap with already detected real clumps. Before injecting the simulated clumps into the six single \textit{ugrizy}-filter band images of the galaxies, we convolved each clump with the effective PSF (ePSF) specific to each host galaxy image and filter band.

We then applied our FRCNN model on each of the 13,789 galaxies with injected simulated clumps. We also applied the same postprocessing steps and extracted the flux peaks from the model detections as described above. A successful detection of a simulated clump is counted if the distance between the extracted flux peak of a predicted clump candidate and simulated clump is less than 0.75 of the image-specific \textit{u}-band seeing FWHM. Our completeness measurements in Section \ref{sec:clumpy_fraction_incomplete}, Figure \ref{fig:sim_clumps_det_completeness_mass_z} and our incompleteness correction of the clumpy fraction (Appendix \ref{sec:clumpy_fraction_incomplete_calc}) are based on the ratio of detected simulated clumps to the total number of simulated clumps. We also used the set of simulated clumps to validate our photometry measurements (Section \ref{sec:clump_phot_val}).

\section{Galaxy segmentation map}\label{sec:hsc_data_gal_extent}
In this section, we briefly describe our method of generating the segmentation masks of the host galaxies \citep[see also][and repeated for convenience]{Popp2026b} that are used to exclude clump detections which are located outside the galaxy extent and to determine the effective radius of the galaxy (Appendix \ref{sec:data_preprocess_reff}). The galaxy segmentation map was created in multiple steps on the \textit{r}-filter band image of the galaxy and following a modified version of the approach used by \citet{Galametz2013} and \citet{Sazonova2021}. In each of the following steps we used the image segmentation routines \texttt{detect\_sources} and \texttt{deblend\_sources} available from the Python library \textsc{Photutils} \citep{LarryBradley2025} with different parameters:
\begin{enumerate}
    \item `hot' mode step:
    \begin{description}
        \item \texttt{data} = science image convolved with a 2-dimensional Tophat filter kernel with a radius of 5 pixels
        \item \texttt{threshold} = 97th percentile value of all pixel values in the science image
        \item \texttt{npixels} = 1 pixel
        \item \texttt{mask} = circular mask with a radius $\mathrm{r}=R_{P,90\%}$
    \end{description}
    \item `cool' mode step:
    \begin{description}
        \item \texttt{data} = science image convolved with a 2-dimensional Tophat filter kernel with a radius of 5 pixels
        \item \texttt{threshold} = standard deviation of all pixel values in the science image
        \item \texttt{npixels} = $1\,\mathrm{arcsec}^2$
        \item \texttt{nlevels} = 32
        \item \texttt{contrast} = $10^{-6}$
    \end{description}
    \item `cold' mode step:
    \begin{description}
        \item \texttt{data} = science image convolved with a 2-dimensional Tophat filter kernel with a radius of 5 pixels
        \item \texttt{threshold} = standard deviation of all pixel values in the science image
        \item \texttt{npixels} = $0.01\times R_{P,90\%}^2$
        \item \texttt{mask} = object mask from `cool' mode
    \end{description}
\end{enumerate}

In the first step (`hot' mode), bright regions in the cutout image were detected by applying a high detection threshold (\texttt{threshold}) so that only pixels with intensity values above the 97th percentile of all pixel values were selected from the input image (\texttt{data}). The input image is a modified version of the science image that is smoothed using a 2-dimensional and isotropic Tophat filter kernel with a radius of 5 pixels. The minimum area, restricted by the minimum number of connected pixels (\texttt{npixels}), is allowed to be small so that only the peak intensity regions were selected in this step. The detected regions were used to mask bright contaminating objects (e.g. foreground stars). From this mask, we excluded bright regions that fall within a circular mask with a radius equal to the SDSS \textit{r}-band 90\% Petrosian radius ($\mathrm{r}=R_{P,90\%}$) around the centre of the target galaxy to avoid excluding possible clump detections..

We then applied a second segmentation and deblending iteration to the smoothed science image (`cool' mode). The detection threshold was lowered to accept pixels that have intensity values above the mean plus standard deviation of all pixel values in the science image. The detected pixels were required to form extended areas with a size of at least $1\,\mathrm{arcsec}^2$ and were deblended into separate regions using the deblending procedure from \textsc{SExtractor} \citep{Bertin1996} that is implemented by \textsc{Photutils}. The values for \texttt{nlevels} and \texttt{contrast} were chosen to separate areas from the target galaxy that are larger than bright point-like sources but are still blended with low-surface brightness features of the main object.

From the regions detected and deblended during the `cool' mode step, the central largest segment and all segments that are located outside a radius of 1.5 times the SDSS \textit{r}-band 90\% Petrosian radius ($\mathrm{r}=1.5\,R_{P,90\%}$) were discarded but the others were kept. Furthermore, we kept all regions that fully overlap with the regions found during the `hot' mode step. The resulting set of regions are likely those areas in the image cutout that are either bright contaminants or blended areas not connected to the target galaxy.

This set of regions was used to mask the smoothed science image to which a last segment detection iteration was applied (`cold' mode). The area threshold (\texttt{npixels}) was changed to a lower value than for the `cool' mode step to include also smaller sources that were missed during the previous step. An additional deblending iteration of the detected segments was not required as all contaminants were expected to be excluded by the applied mask. We kept only the large central segmentation map that corresponds to the target galaxy's extent from the detected segments. In a final step, we smoothed the outline of the galaxy segmentation map using a majority filter kernel with a footprint equal to 10\% of the image size.

\section{Measuring the effective radius of the galaxies}\label{sec:data_preprocess_reff}
As galaxies vary in physical size, we normalise the distance to the centre by the half-light or effective radius $r_{\mathrm{eff}}$, which is determined from the surface brightness profile of the galaxy. We measured the radially averaged flux of the object using:
\begin{equation}\label{eq:hsc_data_preprocess_reff_surfBrightProfile}
    F_{\mathrm{src}}(R) = 2\pi \int_0^R S(r) r\,\dd r
\end{equation}
where $F_{\mathrm{src}}(R)$ is the total flux of the source at radius $R$ and $S(r)$ is the radially averaged surface brightness at radius $r \leq R$ from the centre of the source.

Following \citet{Petrosian1976}, the average flux $\langle{F_{\mathrm{src}}}(R)\rangle$ is the flux $F_{\mathrm{src}}(R)$ divided by the area of a circle with radius $R$:
\begin{equation}\label{eq:hsc_data_preprocess_reff_avgIntensity}
    \langle{F_{\mathrm{src}}}(R)\rangle = \frac{2\pi \int_0^R S(r) r\,\dd r}{\pi R^2}.
\end{equation}
The Petrosian index $\eta (R)$ is defined as the ratio of the average flux at $R$ and the flux at that radius:
\begin{equation}\label{eq:hsc_data_preprocess_reff_petroIndex}
    \eta(R) = \frac{\langle{F_{\mathrm{src}}}(R)\rangle}{F_{\mathrm{src}}(R)} = \frac{2\pi \int_0^R S(r) r\,\dd r}{\pi R^2\, F_{\mathrm{src}}(R)}.
\end{equation}
From $\eta (R)$ the Petrosian radius $R_P$ can be defined as the radius $R$, for which the integrated flux equals the average flux divided by $\eta(R_P)$:
\begin{equation}\label{eq:hsc_data_preprocess_reff_petroRadius}
    F_{\mathrm{src}}(R_P) = \frac{\langle{F_{\mathrm{src}}}(R_P)\rangle}{\eta(R_P)} = \frac{1}{\eta(R_P)}\frac{2\pi \int_0^{R_P} S(r) r\,\dd r}{\pi R_P^2}.
\end{equation}
The inverse of the Petrosian index has often been set to $1/\eta(R_P)=0.2$ as a practical compromise between seeing variations and signal to noise ratio \citep[SNR or $S/N$, e.g. SDSS,][]{York2000,Yasuda2001,Shimasaku2001}. We also used a value of $0.2$ to define the Petrosian radius $R_P$, so that the total Petrosian flux $F_{\mathrm{petro}}$ for every source galaxy is then given by Equation \ref{eq:hsc_data_preprocess_reff_surfBrightProfile} to:
\begin{equation}\label{eq:hsc_data_preprocess_reff_petroFlux}
    F_{\mathrm{petro}} = 2\pi \int_0^{kR_P} S(r) r\,\dd r,
\end{equation}
where $k=2.0$ has been chosen by SDSS for the same practical reasons that informed our choice of $1/\eta(R_P)$.

Finally, the half-light radius $r_{\mathrm{eff}}$ is then the radius $R$ for which:
\begin{equation}\label{eq:hsc_data_preprocess_reff_petroHalfFlux}
    F_{\mathrm{src}}(r_{\mathrm{eff}}) = \frac{1}{2}F_{\mathrm{petro}}.
\end{equation}

We determined $r_{\mathrm{eff}}$ for every galaxy with valid image data in the \textit{gri}-filter bands using the Python package \textsc{PetroFit} \citep{Geda2022}. The average radial surface brightness was measured using elliptical apertures that were centred on the point of maximum flux of the source object. The flux-weighted centroid of the source object was determined by the HSC pipeline and its coordinates are available from the HSC-SSP PDR3 catalogue \citep{Bosch2017}. The area that lies outside the host galaxy extent (Appendix \ref{sec:hsc_data_gal_extent}) was masked to reduce contamination from neighbouring objects that do not belong to the source object \citep[see][]{Popp2026b}. \textsc{Photutils} applies the \textsc{SExtractor} \citep{Bertin1996} centroid and morphological parameters functions that provide elongation and position angles for the source object. The same parameters were used to define the elliptical apertures with radii $r_i$ and $\Delta r=r_{i+1} - r_i = 1\,\mathrm{px}$ to a maximum radius $r_{\mathrm{max}}$ of half the image cutout size in the $x-$direction.

Shape measurements of galaxies were done for HSC-SSP using the \textit{i}-band image as the primary source with the best SNR \citep{Aihara2017}. If the shape measurement failed or the \textit{i}-band image did not pass the photometric quality flags, the \textit{r}-band image was used instead \citep{Mandelbaum2017}. We applied a similar approach to our measurements using primarily the \textit{i}-band image and then the \textit{r}-band and \textit{g}-band images as fallback. From the measurement of the half-light radius, semi-major/semi-minor axes, the elongation (i.e. the ratio of the semi-major to semi-minor axis lengths) and position angles for the source objects were also obtained.

\section{Incompleteness correction}\label{sec:clumpy_fraction_incomplete_calc}
In this appendix we describe the mathematical process that we used to correct the fraction of clumpy galaxies $f_{\mathrm{clumpy}}$ due to potential clumps that were not identified by our clump detector. 

The parameters that primarily impact the detection performance of the FRCNN model, and over which the completeness of the model detections varies significantly, are the \textit{u}-band flux (Fig. \ref{fig:hsc_phys_props_galaxy_completeness_clumps_a}), the colour (\textit{u}-\textit{r}) (Fig. \ref{fig:hsc_phys_props_galaxy_completeness_clumps_b}) and the contrast between the clump and host galaxy in the \textit{u}-band (Fig. \ref{fig:hsc_phys_props_galaxy_completeness_clumps_c}). The intrinsic clump properties (i.e. stellar mass) and host galaxy properties do not directly impact the detection performance but they are correlated with the flux, colour and contrast measured for each clump.

\begin{figure*}
    \centering
    \subfloat[\centering Count of observed clumps. \label{fig:hsc_phys_props_galaxy_completeness_bins_real_counts}]{{\includegraphics[width=1.0\textwidth]{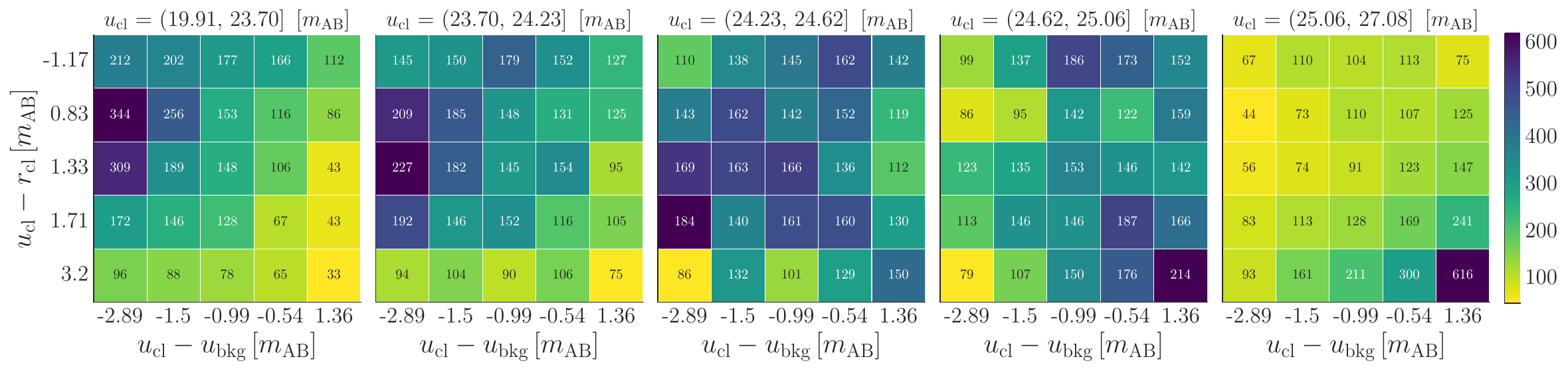} }}
    \\
    \subfloat[\centering Estimated detection completeness of the simulated clumps. \label{fig:hsc_phys_props_galaxy_completeness_bins}]{{\includegraphics[width=1.0\textwidth]{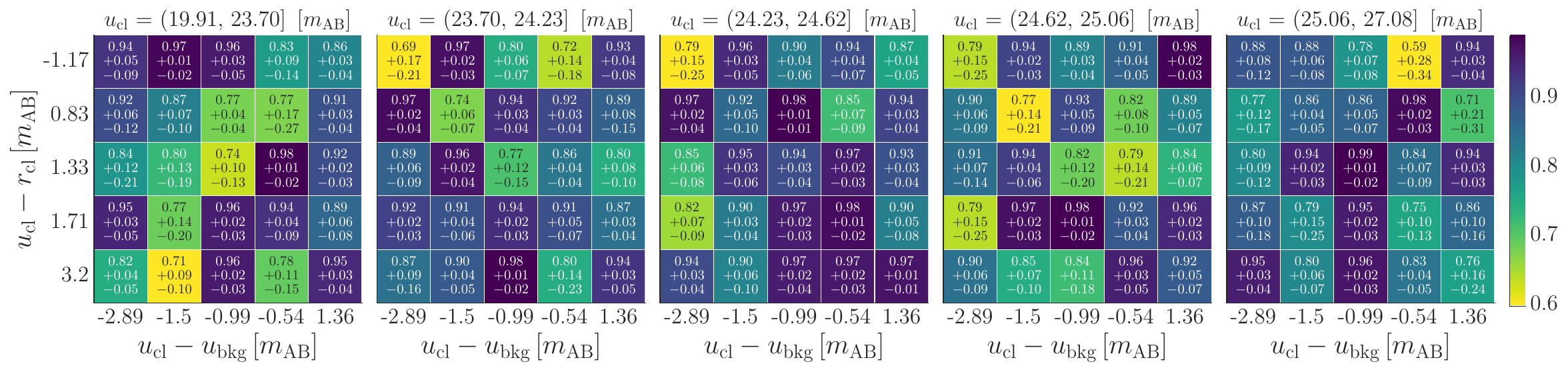} }}
    \caption[Count of observed clumps and estimated detection completeness of the simulated clumps binned over \textit{u}-band flux, contrast and colour (\textit{u}-\textit{r}).]{Count of observed clumps and estimated detection completeness of the simulated clumps binned over a $5\times 5\times 5$ grid spanning the \textit{u}-band flux $u_{\mathrm{cl}}$, \textit{u}-band contrast between the clump and underlying diffuse galaxy background $u_{\mathrm{cl}}-u_{\mathrm{bkg}}$ and colour $u_{\mathrm{cl}}-r_{\mathrm{cl}}$ ranges of the observed clumps. The shown values within the bins are the count of observed clumps (a) and the estimated detection completeness with the 16th and 84th percentiles (b).}
    \label{fig:hsc_phys_props_galaxy_completeness_bins_real_counts_fraction}
\end{figure*}

We first partitioned the observed clumps with a clump-galaxy \textit{u}-band flux ratio of $F_{u,\mathrm{cl}}/F_{u,\mathrm{gal}} \geq 0.08$ into five bins defined over the ranges of the three key parameters \textit{u}-band flux, contrast and colour (\textit{u}-\textit{r}). The bin edges for each parameter were chosen so that the bins contain equal numbers of clumps for that parameter. The resulting 125 bins span a $5\times 5\times 5$ grid. The clump counts in each bin are shown in Figure \ref{fig:hsc_phys_props_galaxy_completeness_bins_real_counts}.

Then, we calculated the recovery fraction in each bin. For the $k$th bin, the recovery fraction $\hat{c}_k=n_{\mathrm{obs},k}/n_{\mathrm{true},k}$ is defined as the ratio of the number of detected simulated clumps $n_{\mathrm{obs},k}$ to the true number of simulated clumps $n_{\mathrm{true},k}$ for bin $k$. 

When considering simulated clumps, for which the value of $n_{\mathrm{true},k}$ is known, then $\hat{c}_k$ provides an estimate of the completeness of the clump detection model for clumps in bin $k$. However, this estimate is vulnerable to large errors if the true number of clumps in bin $k$ is small because small statistical fluctuations in either $n_{\mathrm{obs},k}$ or $n_{\mathrm{true},k}$ can produce large fluctuations in $\hat{c}_k$. To mitigate this, we followed a Bayesian approach as described by \citet{Cameron2011}.

Given the true probability $p_{\mathrm{c},k}$ of the model detecting a clump in bin $k$ (i.e. the true model completeness for bin $k$) and the probability $q_{\mathrm{c},k} = 1-p_{\mathrm{c},k}$ that the model does not detect a clump in bin $k$, the likelihood $p(\hat{c}_k\,|\,p_{\mathrm{c},k})$ of observing $\hat{c}_k$ (or, equivalently, $n_{\mathrm{obs},k}$) for bin $k$ must satisfy:
\begin{equation}
    p(\hat{c}_k\,|\,p_{\mathrm{c},k}) \propto p_{\mathrm{c}, k}^{n_{\mathrm{obs},k}}\,q_{\mathrm{c},k}^{n_{\mathrm{true},k} - n_{\mathrm{obs},k}}.
\end{equation}
\citet{Cameron2011} showed that normalising $p(\hat{c}_k\,|\,p_{\mathrm{c},k})$ in the range $[0,1]$ yields a Beta distribution:
\begin{equation}
    B_{k}(a, b) = \frac{(a+b-1)!}{(a-1)!(b-1)!}p_{\mathrm{c}, k}^{a-1}q_{\mathrm{c}, k}^{b-1}
\end{equation}
where $a = n_{\mathrm{obs},k}+1$ and $b=n_{\mathrm{true},k} - n_{\mathrm{obs},k}+1$. For our simulated clumps, $n_{\mathrm{obs},k}$ and $n_{\mathrm{true},k}$ are both known, so the shape of $B_{k}(a, b)$ can be fully specified.

We assumed that all values of $\hat{c}_k$ are equally likely and defined a uniform prior $p_{\mathrm{prior}}(\hat{c}_k)=1$ for all $\hat{c}_k=[0,1]$. Using Bayes' theorem, the normalised likelihood $p(\hat{c}_k\,|\,p_{\mathrm{c},k})$ is equal to the posterior distribution for $p_{\mathrm{c},k}$:
\begin{equation}
    p(p_{\mathrm{c},k}\,|\,\hat{c}_k) = B_{k}(a, b).
\end{equation}
We use the median of $B_{k}(a, b)$ to provide a robust estimate for $p_{\mathrm{c},k}$ and use the 16th and 84th percentiles of $B_{k}(a, b)$ to quantify the uncertainty on the estimate for $p_{\mathrm{c},k}$ in each bin $k$ (Fig. \ref{fig:hsc_phys_props_galaxy_completeness_bins}). 

Then, for the real clumps $n_{\mathrm{obs,real},k}$ that were detected by the model in our sample of galaxies and have observable properties that place them in bin $k$, the true number of real clumps $n_{\mathrm{true,real},k}$ in bin $k$ can be estimated using:
\begin{equation}
    n_{\mathrm{true,real},k} = \frac{n_{\mathrm{obs,real},k}}{p_{\mathrm{c},k}}.
\end{equation}
A population of real galaxies contains $n_{\mathrm{obs,pop}}$ clump detections with properties that place them in a variety of different bins $k_i$, where $i=\{1, 2, \ldots ,n_{\mathrm{obs,pop}}\}$. The true number of clumps $n_{\mathrm{true,pop}}$ in that population can be estimated by summing $1/p_{\mathrm{c},k_i}$ over all detected clumps:
\begin{equation}
    n_{\mathrm{true,pop}} = \sum_{i=1}^{n_{\mathrm{obs,pop}}} \frac{1}{p_{\mathrm{c},k_i}}
\end{equation}
and the estimated fraction of recovered clumps $\hat{c}_{\mathrm{pop}}$ for the population is then:
\begin{equation}\label{eq:hsc_phys_props_galaxy_compl_clumps_per_gal_completeness}
    \hat{c}_{\mathrm{pop}} = \frac{n_{\mathrm{obs,pop}}}{n_{\mathrm{true,pop}}} = \frac{n_{\mathrm{obs,pop}}}{\displaystyle\sum_{i=1}^{n_{\mathrm{obs,pop}}} \frac{1}{p_{\mathrm{c},k_i}}}.
\end{equation}

For this population of galaxies, we assumed that the true number of clumps per galaxy follows an exponential distribution, such that the probability that any single galaxy contains $n^{*}_{\mathrm{true}}$ clumps:
\begin{equation}\label{eq:hsc_phys_props_galaxy_compl_clumps_per_gal}
    p(n^{*}_{\mathrm{true}}\,|\,\gamma) \sim e^{-\gamma n^{*}_{\mathrm{true}}},
\end{equation}
where $\gamma$ is an unknown rate parameter.

To estimate $p(n^{*}_{\mathrm{true}}\,|\,\gamma)$, we started with a first guess of the parameter $\gamma$ and sampled $10^5$ values of $n^{*}_{\mathrm{true}}$ from $p(n^{*}_{\mathrm{true}}\,|\,\gamma)$ with each value representing the population of clumps in a single galaxy. We discarded a fraction of $1-\hat{c}_{\mathrm{pop}}$ galaxies as well as those galaxies that were modelled with no clumps. This yields a new distribution of clumps per galaxy that is a prediction for what the observed distribution of clumps per galaxy would be \textit{if} the true value of $\gamma$ was equal to our initial guess (i.e. the ``incomplete'' distribution). We then compared the mean number of clumps per modelled galaxy to the observed mean number of clumps per galaxy and continued this process, using gradient descent to adjust $\gamma$ in each step for a new sample of $10^5$ modelled galaxies, until both means differed by $<0.001$. Once the ``best fit'' value of $\gamma$ was found, it can be used in combination with Equation \ref{eq:hsc_phys_props_galaxy_compl_clumps_per_gal} to infer the true, incompleteness-corrected, distribution for $n^{*}_{\mathrm{true}}$.

\begin{figure}
    \centering
    \includegraphics[width=0.7\columnwidth]{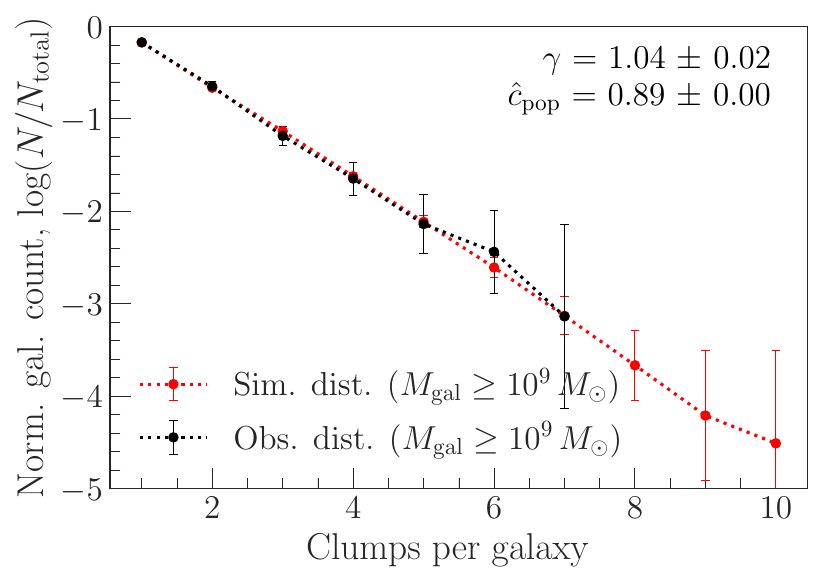}
    \caption[Observed and modelled ``incomplete'' distribution of clumps per galaxy.]{Observed and modelled ``incomplete'' distribution of clumps per galaxy for the mass-complete galaxy sample and clumps that have $F_{u,\mathrm{cl}}/F_{u,\mathrm{gal}} \geq 0.08$. The observed distribution is shown in black and the modelled distributions in red. Error bars show the standard errors. The ``best fit'' value for $\gamma$ and the measured recovery fraction $\hat{c}_{\mathrm{pop}}$ are also shown in the plots as annotations.}
    \label{fig:hsc_phys_props_galaxy_compl_clumps_per_gal}
\end{figure}

In Figure \ref{fig:hsc_phys_props_galaxy_compl_clumps_per_gal}, we show the observed number of galaxies and the final modelled ``incomplete'' distributions as a function of the number of clumps per galaxy for clumps with $F_{u,\mathrm{cl}}/F_{u,\mathrm{gal}} \geq 0.08$. A separate estimate of $\gamma$ is made to estimate the incompleteness-corrected distribution of $n^{*}_{\mathrm{true}}$ in each of the galaxy sample splits that were used in our analysis.

With estimates for $\hat{c}_{\mathrm{pop}}$ and $p(n^{*}_{\mathrm{true}}\,|\,\gamma)$, the probability $p_{\mathrm{miss}}$ that a galaxy is a false negative, i.e. it is a true clumpy galaxy but no clumps above the clump-galaxy flux ratio were detected, is given by:
\begin{equation}
    p_{\mathrm{miss}} = \sum_{n^{*}_{\mathrm{true}}=1}^{\infty} p(n^{*}_{\mathrm{true}}\,|\,\gamma) (1-\hat{c}_{\mathrm{pop}})^{n^{*}_{\mathrm{true}}}.
\end{equation}

The false negative probability $p_{\mathrm{miss}}$ can finally be used to correct the observed clumpy fraction $\widehat{f}_{\mathrm{clumpy}}$ to obtain an estimate for the true clumpy fraction $f_{\mathrm{clumpy}}$:
\begin{equation}
    f_{\mathrm{clumpy}} = \frac{1}{1-p_{\mathrm{miss}}}\widehat{f}_{\mathrm{clumpy}}.
\end{equation}

We used the same approach to correct the observed clumpy fraction for a clump definition based on a clump stellar mass threshold of $M_{\mathrm{cl}} \geq 10^7\,M_\odot$ instead of the clump-galaxy flux ratio. However,  in this case we added a fourth parameter, the clump stellar mass $\log(M_{\mathrm{cl}}/M_\odot)$, to our grid of bins because the detection performance of the FRCNN model decreases significantly with decreasing stellar mass of the clumps (Fig. \ref{fig:hsc_phys_props_galaxy_completeness_clumps_e}). We partitioned the observed clumps into five stellar mass bins and chose bin edges so that the each stellar mass bin contains roughly equal number of clumps. Our new grid contains 625 bins, for which we calculate the recovery fraction $\hat{c}_k$ in the same way as described above.


\bsp	
\label{lastpage}
\end{document}